\PassOptionsToPackage{authoryear,longnamesfirst}{natbib}
\documentclass[a4paper,fleqn]{cas-dc}

\usepackage{algorithmic}
\usepackage{algorithm}
\usepackage[authoryear]{natbib}
\usepackage[authoryear,longnamesfirst]{natbib}

\def\tsc#1{\csdef{#1}{\textsc{\lowercase{#1}}\xspace}}
\tsc{WGM}
\tsc{QE}
\newcommand{\citeneed}{}
\newcommand{\norm}[1]{\left\lVert#1\right\rVert}
\DeclareMathOperator*{\argmin}{argmin}

\usepackage{pythonhighlight}
\hypersetup{ % play with the different link colors here
	colorlinks,
	citecolor=cyan,
	filecolor=blue,
	linkcolor=cyan,
	urlcolor=cyan % set to black to prevent printing blue links
}
\begin{document}
\let\WriteBookmarks\relax
\def\floatpagepagefraction{1}
\def\textpagefraction{.001}

%\shorttitle{<short title of the paper for running head>}    

%\shortauthors{<short author list for running head>}  

% \title [mode = title]{<main title>}  

\title[mode = title]{Design and Implementation of Real-Time Localization System (RTLS) based on UWB and TDoA Algorithm}

%\tnotemark[1,2]

%\tnotetext[1]{This document is the results of the %research project funded by the National Science %Foundation.}

%\tnotetext[2]{The second title footnote which is a %longer text matter to fill through the whole text %width and overflow into another line in the footnotes %area of the first page.}


\author[1]{Fengyun Zhang}
[
    style=chinese,
    %type=editor,
    auid=001,
    bioid=1,
    %prefix=Sir, 
    %role=Researcher, 
    orcid=0000-0001-8274-0616
]
\ead{zhangfy2019@mail.sustech.edu.cn} 
% Credit authorship of first author
\credit{Conceptualization, Methodology, Writing Original draft, Writing-Reviewing and Editing}
%\cormark[1]


%\ead[url]{www.cvr.cc,www.tug.org.in}


\address[1]{Department of Computer Science and Engineering, Southern University of Science and Technology, Shenzhen, China}

%\affiliation[<aff no>]{organization={}, addressline={}, city={},postcode={}, state={},country={}}

\author[2]{Li Yang}
[
    style=chinese,
    %type=editor,
    auid=002,
    bioid=2,
    %prefix=Sir, 
    %role=Researcher, 
    orcid=0000-0001-5017-2095
]
% Email id of the second author
\ead{yangli.aeup@gmail.com} 
% Credit authorship of second author
\credit{Conceptualization, Funding acquisition}
% Address
\address[2]{Army Engineering University of PLA, Nanjing, China}

\author[1]{Yuhuan Liu}
[
    style=chinese,
    %type=editor,
    auid=003,
    bioid=3,
    %prefix=Sir, 
    %role=Researcher, 
    orcid=0000-0002-7507-4579
]
% Email id of the third author
\ead{liuyh2019@mail.sustech.edu.cn} 
% Credit authorship of third author
\credit{Methodology, Writing-Reviewing and Editing}

\author[1, 3]{Yulong Ding}
[
    style=chinese,
    %type=editor,
    auid=004,
    bioid=4,
    %prefix=Sir, 
    %role=Researcher, 
    orcid=0000-0002-7159-5107
]
% Email id of the fouth author
\ead{dingyl@sustech.edu.cn} 
% Credit authorship of fouth author
\credit{Supervision,Conceptualization, Writing-Reviewing and Editing}
% Address
\address[3]{Academy for Advanced Interdisciplinary Studies, Southern University of Science and Technology, Shenzhen, China}

\author[1, 4]{Shuang-Hua Yang}
[
    style=chinese,
    %type=editor,
    auid=005,
    bioid=5,
    %prefix=Sir, 
    %role=Researcher, 
    orcid=0000-0003-0717-5009
]
% Email id of the fouth author
\ead{yangsh@sustech.edu.cn} 
% Credit authorship of fouth author
\credit{Methodology, Supervision, Conceptualization, Writing-Reviewing and Editing, Funding acquisition}
% Address
\address[4]{PCL Research Center of Networks and Communications, Peng Cheng Laboratory, Shenzhen, China}
\cormark[1] 

\author[5]{Hao Li}
[
    style=chinese,
    %type=editor,
    auid=006,
    bioid=6,
    %prefix=Sir, 
    %role=Researcher, 
    orcid=0000-0002-0253-5733
]
% Email id of the fouth author
\ead{lihao.stndl@gmail.com} 
% Credit authorship of fouth author
\credit{Conceptualization, Funding acquisition}
% Address
\address[5]{Science and Technology on Near-surface Detection Laboratory, Wuxi, China}

\cormark[1] 

%\affiliation[<aff no>]{organization={},addressline={}, city={},postcode={}, state={},country={}}

% \cortext[1]{Corresponding author}
\cortext[1]{Corresponding author}

%\cortext[2]{Principal corresponding author} 

%\fntext[1]{This is the first author footnote. but is common to third author as well.}

%\fntext[2]{Another author footnote, this is a very long footnote and it should be a really long footnote. But this footnote is not yet sufficiently long enough to make two lines of footnote text.}

%\fntext[3]{K. Berry is the editor of \TeX Live.}

%\nonumnote{This note has no numbers. In this work we demonstrate $a_b$ the formation Y\_1 of a new type of polariton on the interface between a cuprous oxide slab and a polystyrene micro-sphere placed on the slab. The evanescent field of the resonant whispering gallery mode (\WGM) of the micro sphere has a substantial gradient, and therefore effectively couples with the quadrupole $1S$ excitons in cuprous oxide.} 


\begin{abstract}[S U M M A R Y]
  Nowadays, accurate localization plays an essential role in many fields, like target tracking and path planning. The challenges of indoor localization include inadequate localization accuracy, unreasonable anchor deployment in complex scenarios, lack of stability, and the high cost. So the universal positioning technologies cannot meet the real application requirements scarcely. To overcome these shortcomings, a comprehensive Ultra Wide-Band (UWB) based RTLS is presented in this paper. We first introduce the architecture of real-time localization system, then propose a new wireless clock synchronization (WCS) scheme, finally discuss the time difference of arrival (TDoA) algorithm. We define the time-base selection strategy for TDoA algorithm, and analyze the relationship between anchor deployment and positioning accuracy. The Extended Kalman Filter (EKF) method is presented for non-linear dynamic localization estimation, and it performs well in terms of stability and accuracy in moving target.
\end{abstract}


\begin{highlights}
\item A better performance UWB based real-time localization system is designed. 
\item A new wireless clock synchronization scheme covering both a single master anchor and multiple master anchors is proposed. 
\item The typical anchor deployment schemes are presented, and the time-base selection strategy for TDoA algorithm are defined.
\end{highlights}

% Keywords

\maketitle

% Main text
\section{Inroduction}\label{Indro}
accuracy position information of person or device is vital for military, security, and commercial applications. For example, knowing the location information of living creature in danger situation can help firefighters with an emergency rescue, and indoor location technology can also facilitate consumers' shopping in the supermarket. Indoor Real-Time Localization System (RTLS) has not yet been widely deployed, although many neoteric technologies, such as computer vision and wireless communications solutions, are adopted \citep{DBLP:journals/sensors/AlarifiAAAAAA16, Alwan2017GradientDL}. It is not easy to integrate many other technologies into a unified solution. With the increase of indoor positioning demand, how to obtain accurate location information becomes particularly important \citep{PUGHAT201719}.
	
There are a few kinds of wireless sensor technologies used in indoor positioning, such as WiFi, RFID \citep{PAPAPOSTOLOU2011902} and BLE. Whereas, high frequency positioning with WiFi needs a high power consumption, so a low-power consumption but with a high localization accuracy solution is needed \citep{DBLP:journals/ijnm/CurranF08}. UWB (Ultra Wide-Band) is a radio technology that uses pulse rather than the carrier to transmit data, ensuring its low-power consumption. The Federal Communications Commission (FCC) has identified that UWB pulses should occupy a broad frequency bandwidth ($>500 MHz$) or a relative bandwidth ($>20\%$) with a restricted frequency band from 3.1 GHz to 10.6 GHz and -41.3 dBm/MHz power density \citep{FederalRegister}. \citep{LI20111894} proposes a low-complexity and noncoherent detector to detect UWB signals in the wireless sensor networks (WSNs). According to the manufacturer's datasheets, indoor point to point measurement using UWB has high accuracy, achieving the accuracy within 10cm. Besides, the frequency band of UWB makes the UWB devices data transmission rate up to 500Mbit/s \citep{DW1000UserManual}, \citep{DW1000Datasheet}. An OpenSource hardware-platform based on the DW1000 UWB chip called Wi-PoS is proposed in \citep{DBLP:journals/sensors/HerbruggenJRRMB19}. The PolyPoint project presents a multi-antenna plan to eliminate the influence of polarization mismatch between anchors in \citep{10.1145/2799650.2799651}. \citep{5606272} proposes and analyzes the UWB-WBAN system, which could be applied to design for WBAN applications. \citep{DBLP:journals/tce/YouL11} proposes an improved pilot-based algorithm based on CFO and SFO in UWB-OFDM systems with cyclic delay diversity. \citep{9409138} studies the employment of UWB in a factory and a Bayesian filtering solution is developed to track the targets. \citep{DBLP:journals/tim/BottiglieroMSM21} introduces a new low-cost RTLS without time synchronization among sensors and uses a one-way communication solution to reduce the consumption of tags. \citep{PELKA201675} presents a unified architecture for location systems to integrate hardware, software and algorithms. \citep{PEASE201798} proposes a semantic IIoT architecture using a communication economical RSSI/ToF ranging method.
	
This paper presents an UWB-based RTLS, including system architecture, wireless clock synchronization scheme, anchor deployment scheme, and time-base selection strategy. The findings made in this paper offer a solid foundation for all available UWB based indoor localization systems design and deployment. The main contributions of this paper can be summarized as follows: 

\begin{enumerate}[1.]	
\item We design a better performance UWB based real-time localization system. According to the characteristics of UWB and TDoA positioning scheme, the system architecture of RTLS is summarized. 
\item We propose a new wireless clock synchronization scheme covering both a single master anchor and multiple master anchors. 
\item We present the typical deployment schemes of a single master anchor and multiple master anchors based on the principle of anchor deployment, define the time-base selection strategy for TDoA algorithm in signal master anchor and multiple master anchors systems, and reveal the relationship between anchor deployment and positioning accuracy.
\end{enumerate}

The rest of the paper is organized as follows. In Section \textbf{2}, a literature review is given. In Section \textbf{3}, a detailed description of RTLS is given, including the architecture of real-time localization system, and the UWB-based Wireless Positioning Network (U-WPN). In Section \textbf{4}, we introduce the Wireless Clock Synchronization (WCS) scheme. The localization algorithm based on TDoA and EKF will be discussed in Section \textbf{5}. The reference scheme of anchor deployment and time-base selection strategy will be given in Section \textbf{6}. Experiments and performance analysis are discussed in Section \textbf{7}, and finally, conclusions and future work are given in Section \textbf{8}. 

\section{Related Works}\label{Rela}
UWB technology has drawn enough attention to outdoor/indoor localization in recent years. Several methods are used for localization in wireless networks \citep{DBLP:conf/cw/Al-AmmarAAAAAA14}, and these approaches can generally be divided into four categories: (RSSI), (ToF), (ToA), and (TDoA). The positioning method based on TDoA has attracted more attention. 
	
The Atlas \citep{DBLP:conf/ipin/TiemannEW16a} realizes a UWB-based project with DWM1000 \cite{b8}. \citep{7492077} focuses on TDoA based WCS techniques, which relies on pairs of packets and a recorded timestamp, is implemented in RTLS. \citep{8533796} proposes an effective synchronization method of the unilateral TDoA applied in ultra-wideband (UWB) localization systems. \citep{8766539} describes an architecture of multi-level IoT positioning system to reduce the deployment cost. \citep{8902100} presents an E-DTDOA based ranging algorithm used for clock drift estimation, which achieves high time resolution. \citep{9098946} investigates multiple clock-drift correction methods for ToA and TDoA, in particular the DW1000 transceiver. A hybrid positioning algorithm that combines ToA and received signal strength (RSSI) measurements are presented in \citep{khan2013hybrid}. \citep{DBLP:journals/tim/ComunielloMA20} proposes a best linear unbiased estimator (BLUE) algorithm based on ultrasound TDoA measurements and investigates the geometrical dilution of precision (GDOP). An algorithm framework that integrates EKF, UKF, and PF is developed in \citep{8939454}. \citep{7818751} describes an UWB based localization system using the TDoA technology. \citep{XU2021102913} proposes an error-ellipse-resampling particle flter method for cooperative target tracking. A sensor network and a hybrid algorithm for tracking based on both RSS and TDoA is presented in \citep{b7014849}.
	
The positioning solutions mentioned above are algorithm oriented and do not fully consider the relationships between hardware, software, localization scheme, and anchor deployment. The RTLS proposed in this paper focuses on the shortcomings of existing solutions. The outcomes such as the architecture of real-time localization system, the framework of central localization engine, the wireless clock synchronization scheme, and the deployment scheme, could be used as a foundation for all the available UWB based indoor localization solutions.

% Use the style of numbering in square brackets.
% If nothing is used, default style will be taken.
%\begin{enumerate}[a)]
%\item 
%\item 
%\item 
%\end{enumerate}  

%\begin{itemize}
%\item 
%\item 
%\item 
%\end{itemize}  

%\begin{description}
%\item[]
%\item[] 
%\item[] 
%\end{description}  

\section{System Design}\label{Main}
This section describes the details of our new UWB real-time localization system (RTLS), including the architecture of real-time localization system, and the UWB-based Wireless Positioning Network (U-WPN).
\subsection{Architecture of Real-time Localization System}
We have completed the development of embedded software, CLE software and application software. Figure 1 below gives a detailed view of the architecture of real-time localization system. The UWB based RTLS can be regarded as a three-layer architecture, including application layer, CLE framework layer, and the U-WPN layer. UWB anchors and tags are working at the U-WPN layer, and the UWB messages could be transmitted between them through ISO/IEC defined protocol. The CLE has a control function unit, a database, and an algorithm unit, which realizes the function of wireless clock synchronization and location estimation. Mobile apps and Web platforms could interact with the CLE framework layer through the open APIs.
\begin{figure*}
	\centering
	  \includegraphics[width=\textwidth,height=4.5in]{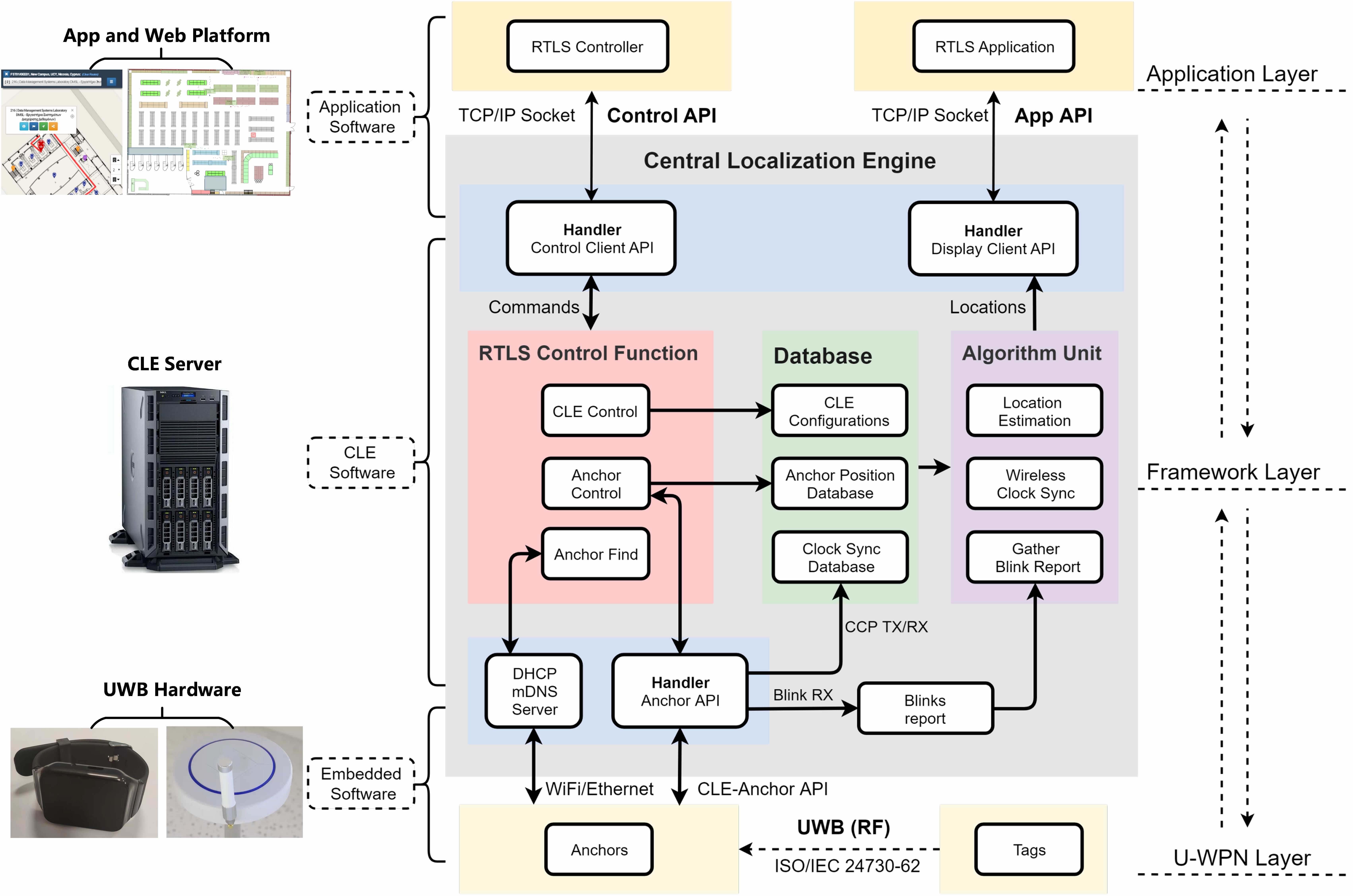}
	\caption{Architecture of Real-Time Localization System.}
	\label{FIG:1}
\end{figure*}

\subsection{UWB-based Wireless Positioning Network}
Figure 2 demonstrates the UWB-based Wireless Positioning Network (\textbf{U-WPN}) , which exchanges the UWB messages and Ethernet communication for a system with four Anchors, one Tag, and one \textbf{CLE} (Central localization engine). MA is the master anchor, and SA2, SA3, SA4 are the slave anchors. T1 is the Tag, and the \textbf{CLE} runs on an upper computer. The Tag transmits a periodic blink frame, which is received and timestamped at the anchors. Each anchor then sends the ToA reports to a Central Localization Engine (\textbf{CLE}), and the \textbf{CLE} uses the ToAs to estimate the Tag's location. Tags periodically send "blink" messages, which are received by all anchors in the range. 
	
To ensure that the ToAs recorded by the anchors are on the same reference clock, we need to eliminate the clock offset and drift of anchors. It is called clock synchronization and is typically achieved via wired clock distribution to the anchors. As an alternative to using a wired timer, the TDoA based RTLS designed in this paper includes a wireless clock synchronization algorithm that employs UWB messages sent between anchors to correct clock drift and offset. 
	
Anchors can be configured as Master anchors (MA) and Slave anchors (SA). MA transmit Clock Calibration Packets (CCP) periodically. Slave anchors receive these CCP and report their reception to the \textbf{CLE} to track the relative clock offset between the sending master anchors and the receiving slave anchors. If an RTLS has more than one master, which could be called multiple masters based RTLS. A "secondary" master can delay sending its CCP by a configured lag time after the reception of a CCP from a "primary" master anchor to prevent from CCP collisions between CCP transmissions. 
	
The purpose of clock synchronization is to record each anchor's clock and calibrate the blink message timestamps to a recorded timestamp. The \textbf{CLE} also performs the TDoA algorithm to estimate the tags' locations. 
% Figure
\begin{figure}
	\centering
		\includegraphics[scale=.055]{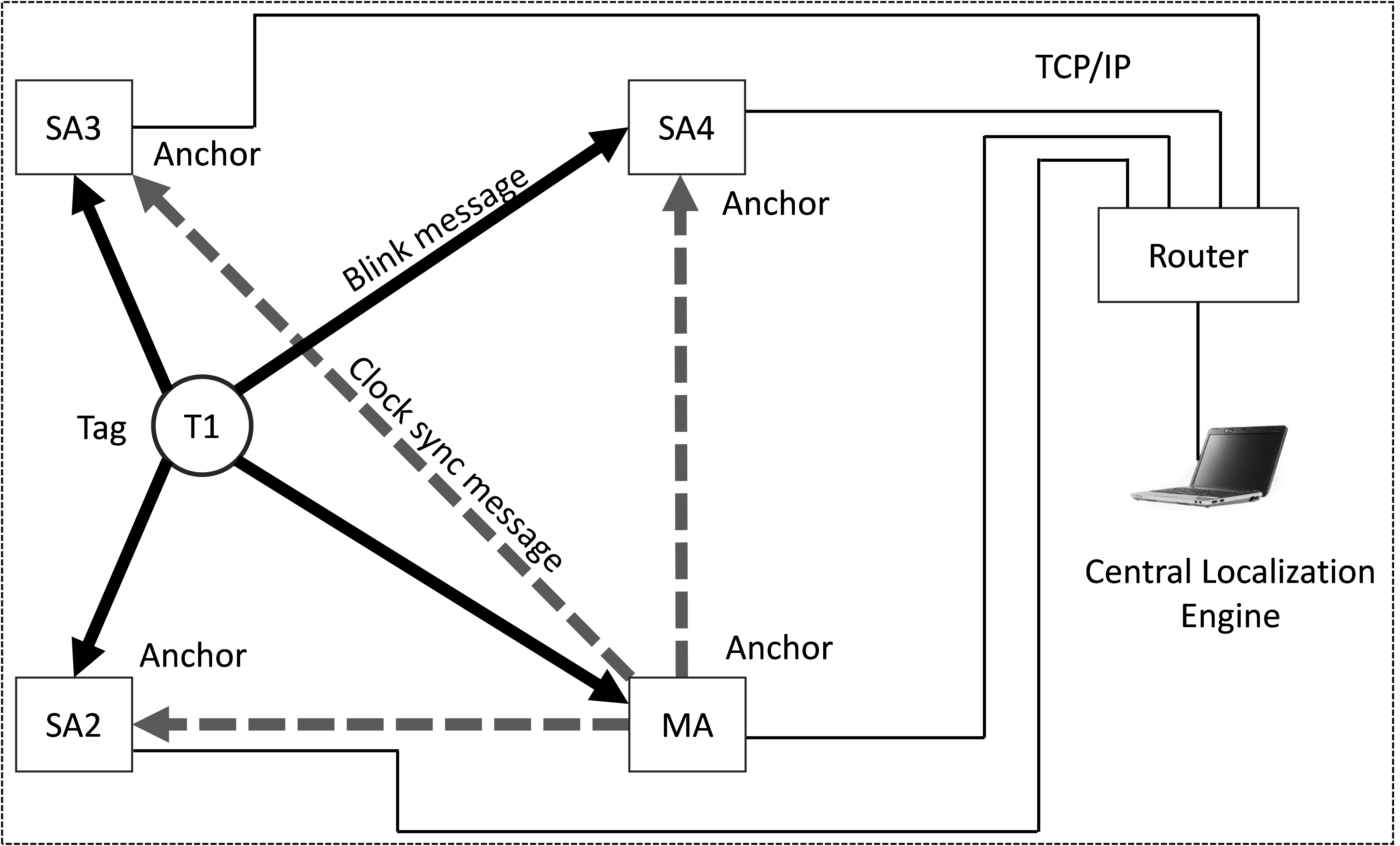}
	\caption{UWB-based Wireless Positioning Network.}
	\label{FIG:2}
\end{figure}

% Uncomment and use as the case may be
%\begin{theorem} 
%\end{theorem}

% Uncomment and use as the case may be
%\begin{lemma} 
%\end{lemma}


\section{Wireless Clock Synchronization}\label{WCS}
Each UWB device equips with a high-resolution timer. The oscillation frequency will drift over time, and we must eliminate the effects of drift by clock calibration. The clock frequency offset and instability significantly impacts positioning accuracy in RTLS. Most of the RTLS utilize time measurement to get ToA or TDoA, which is used for position calculation. All the anchors need to be synchronized, as the precise timestamp is essential for location estimation. Three main issues should be addressed in synchronization \citep{Lee2011NonsynchronisedTD}: 
\begin{itemize}
	\item \textbf{Offset synchronization} – Ensure the recorded timestamp between anchors using the same reference time; 
	\item \textbf{Drift compensation} – Eliminate frequency deviation caused by temperature and other environmental factors. \item \textbf{Antenna delay calibration} – Eliminate the changes of different UWB devices's internal propagation delay.  
\end{itemize}
	
The measurement delay in the timestamp includes transmitting antenna delay and receiving antenna delay. These antenna delays are specifically internal to the chip and have not been included in time of flight (ToF). The solutions proposed in \citep{8483104, 7492077} can be used for antenna delay calibration.

The wired clock synchronization scheme is an universal solution. However, additional clock synchronizing timer and transmission lines add to the difficulty of anchor deployment, so it is not suitable for complex environments. Hence a wireless clock synchronization solution without extra equipment is urgently needed. The method proposed in \citep{7492077} relies on the pair of packets and a known recorded timestamp, which uses the remained nodes' corrected timestamps to carry out WCS. Meanwhile, a simple clock model is used in \citep{DBLP:conf/ipin/TiemannEW16a} for wireless clock correction, based on the offset and the drift. It is necessary to retain clock models for each anchor. \citep{6881189} proposes a clock synchronization scheme that can be used in multilateral systems based on a reference anchor placed at a known fixed location.
	
Clock offset correction is easier to solve when we know each anchor's reference clock's deviation, but the clock drift is not easy to eliminate due to the different clock modules in the anchors. Figure 2 depicts a basic UWB-based Wireless Positioning Network \textbf{U-WPN} in Section \textbf{3}. We will use the \textbf{U-WPN} to introduce our proposed wireless clock synchronization scheme in this section. 
	
Each Anchor and Tag have its own timer, and they are un-synchronized. The exact clock information of another device is unknown, which can only be obtained through a timestamp. Besides, the precise distance between the sending device and the receiving device is unknown. If all the slave anchors could receive the \textbf{CCPs} sent by one master anchor, so we adopt the scheme of \textbf{WCS with a single master}. If the \textbf{CCP} sent by a master anchor cannot be received by all slave anchors, multiple master anchors need to be deployed in the location area. In that case, the \textbf{WCS with multiple masters} need to be adopted for completed coverage.

\subsection{WCS with a single master}
The overall diagram of WCS with a single master is demonstrated in Figure 3, which shows our proposed positioning system receiving and sending positioning and synchronization messages on the timeline. \textbf{T} is tag, \textbf{MA} is a master anchor and \textbf{SA2-SA4} are slave anchors. There are five timelines, with the top one representing the Tag and the bottom four belonging to the individual anchors. The dark dashed lines represent the positioning packet (\textbf{Blinks}) sent by the Tag with a sending period of 1s, while the light dashed lines represent the synchronous packet (\textbf{CCPs}) sent by the master anchor at an interval of 150ms.

Once the Tag sends the \textbf{Blink} or the master anchor sends the \textbf{CCP}, the slave anchors in the corresponding deployment area will receive the \textbf{Blink} or \textbf{CCP} and record the timestamp in the anchor's clock system respectively. For the master anchor, in addition to receiving the Blink from the Tag similar to a slave anchor, it is also necessary to periodically send the \textbf{CCP} and record the sending timestamp.
\begin{figure}
	\centering
		\includegraphics[scale=.06]{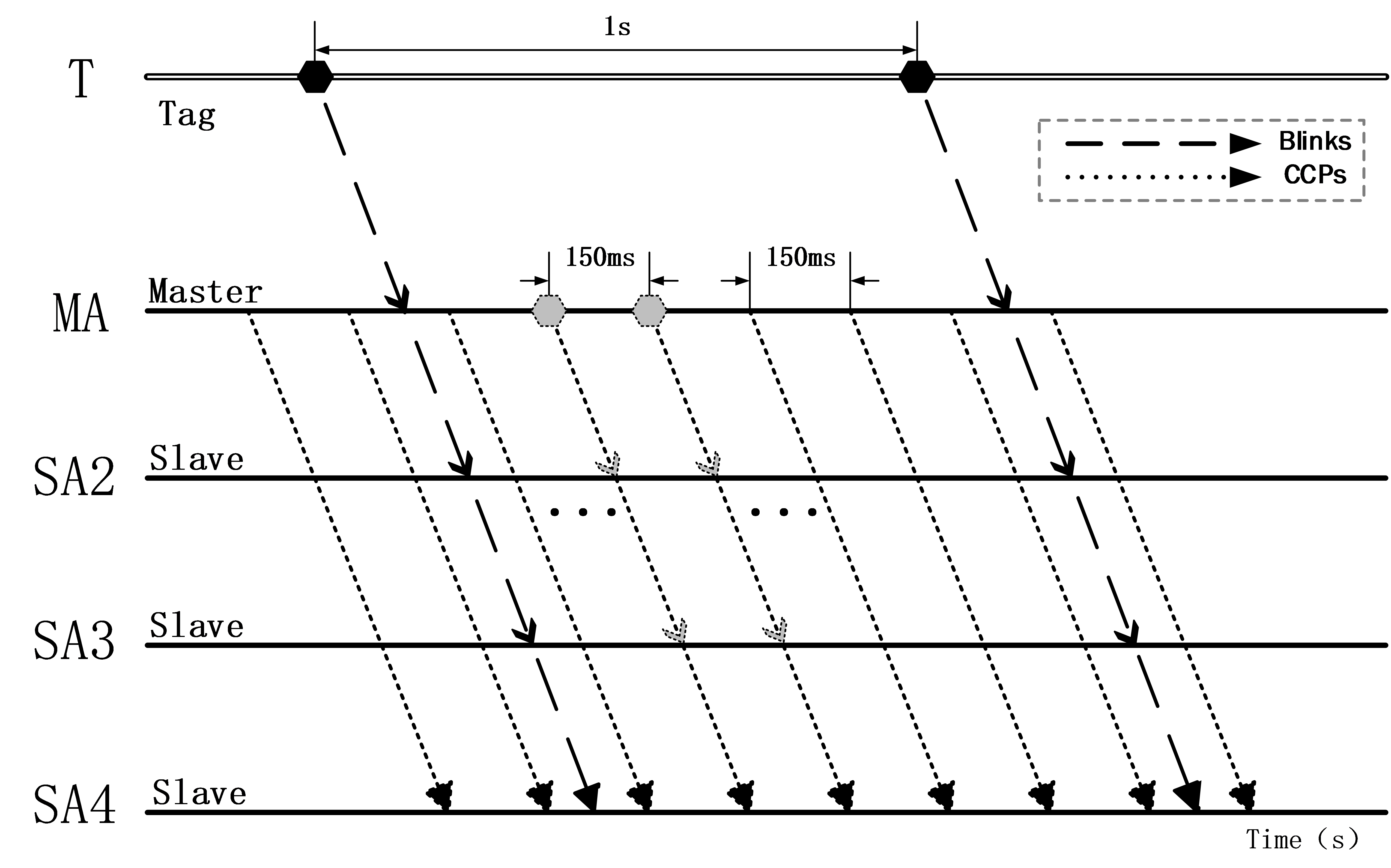}
	\caption{Diagram of  WCS with a single master.}
	\label{FIG:3}
\end{figure}

To illustrate the process of clock synchronization in detail, we use a simple case with one tag (T), one master anchor (MA), and one slave anchor (SA), as shown in Figure 3. After receiving the \textbf{Blink}, \textbf{MA} and \textbf{SA} will record the received timestamp ($R_{x0}$, $R_{x1}$) and serial number (\textbf{SeqNum}) of this \textbf{Blink}. Meanwhile, the \textbf{MA} will record the sending timestamp ($T_{s1}$, $T_{s2}$) and the corresponding serial number (SeqNum1, SeqNum2) when it sends the \textbf{CCP}. When the \textbf{CCP} reaches the \textbf{SA} node, the timestamp ($R_{s1}$, $R_{s2}$) and the corresponding serial number will also be recorded. 

The system clock drift caused by quartz crystal illustrates a certain regularity, so we build a scale coefficient model to correct the clock drift. As shown in Figure 4, an original TDoA can be expressed as: $TDoA_{raw}=R_{x1}-R_{x0}$, and the scale coefficient of calibration can be set as $K$. 
\begin{equation}
	K=\frac{T_{s1}-T_{s2}}{R_{s1}-R_{s2}}
\end{equation}
The corrected TDoA  is $TDoA_{sync}$:
\begin{equation}
	TDoA_{sync}=K*TDoA_{raw}
\end{equation}
where, $R_{x1}$ and $R_{x0}$ are the timestamps recorded by \textbf{SA} and \textbf{MA} when they receive a \textbf{Blink}, $R_{s1}$ and  $R_{s2}$ are the timestamps recorded by \textbf{SA} when it receives the \textbf{CCPs}, $T_{s1}$ and $T_{s2}$ are the timestamps recorded by \textbf{MA} when it sends the \textbf{CCPs}. Finally, the Kalman filters will be adopted to track each anchor’s clock offset after gathering all the \textbf{$TDoA_{sync}s$}. Based on this scheme, the clock of each anchor can be synchronized. 
\begin{figure}
	\centering
		\includegraphics[scale=.1]{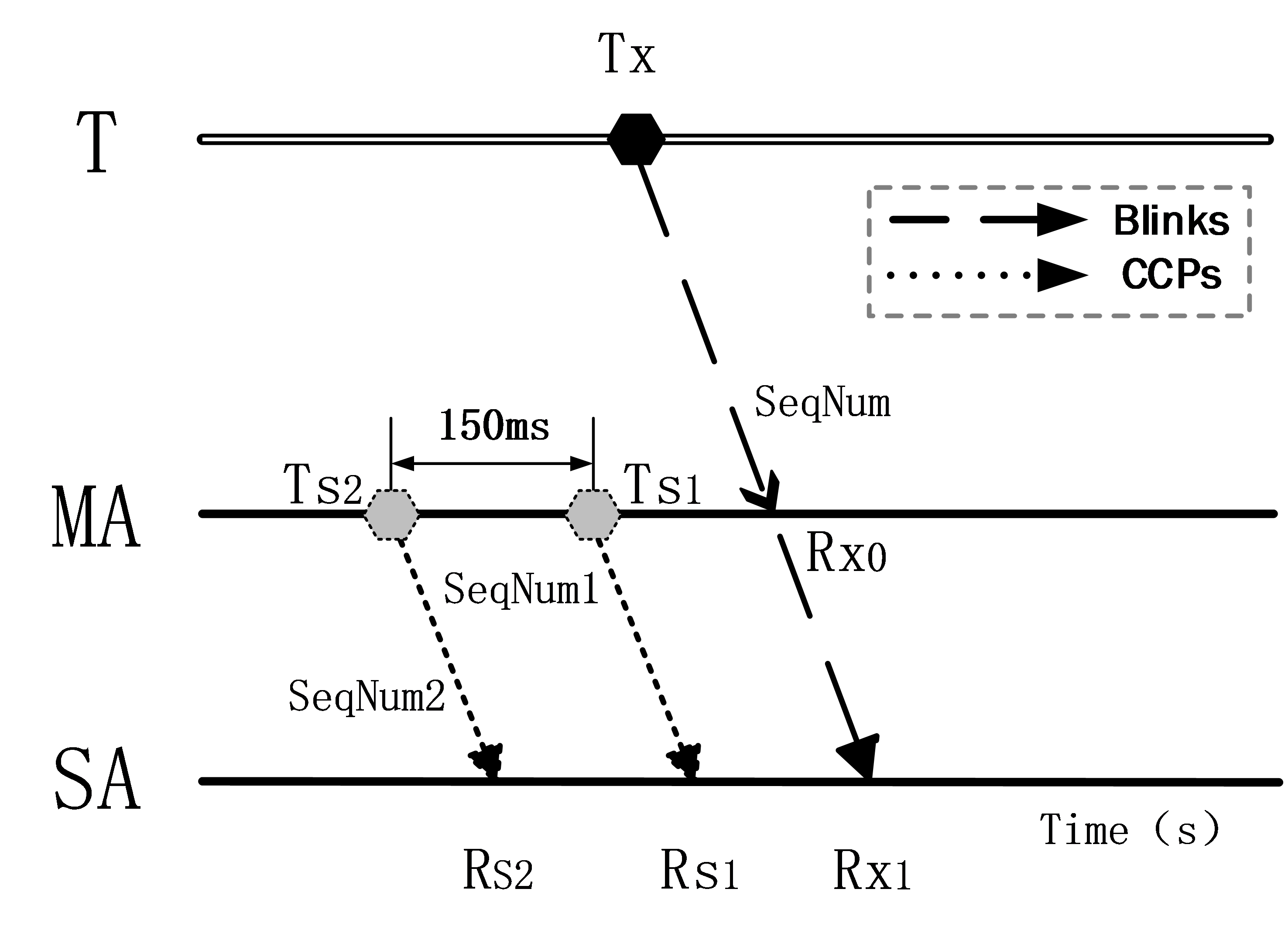}
	\caption{Diagram of the details of WCS.}
	\label{FIG:4}
\end{figure}

\subsection{WCS with multiple masters}
The overall diagram of WCS with multiple masters is demonstrated in Figure 5, including one primary \textbf{MA} (\textbf{MA1}), five secondary \textbf{MAs} (\textbf{MA2-MA6}), and nineteen \textbf{SAs}. The \textbf{CCPs} between anchors (\textbf{MA} and \textbf{SA}, or \textbf{MA} and \textbf{MA}) are in dotted lines. The deployment of \textbf{MAs} has been made to ensure that all secondary master anchors can communicate with the master anchor of the upper level, and any slave anchor can receive the synchronization signal of at least one master anchor. The rest part of this section describes in detail how the WCS scheme with multiple masters works. 
\begin{figure}
	\centering
		\includegraphics[scale=.035]{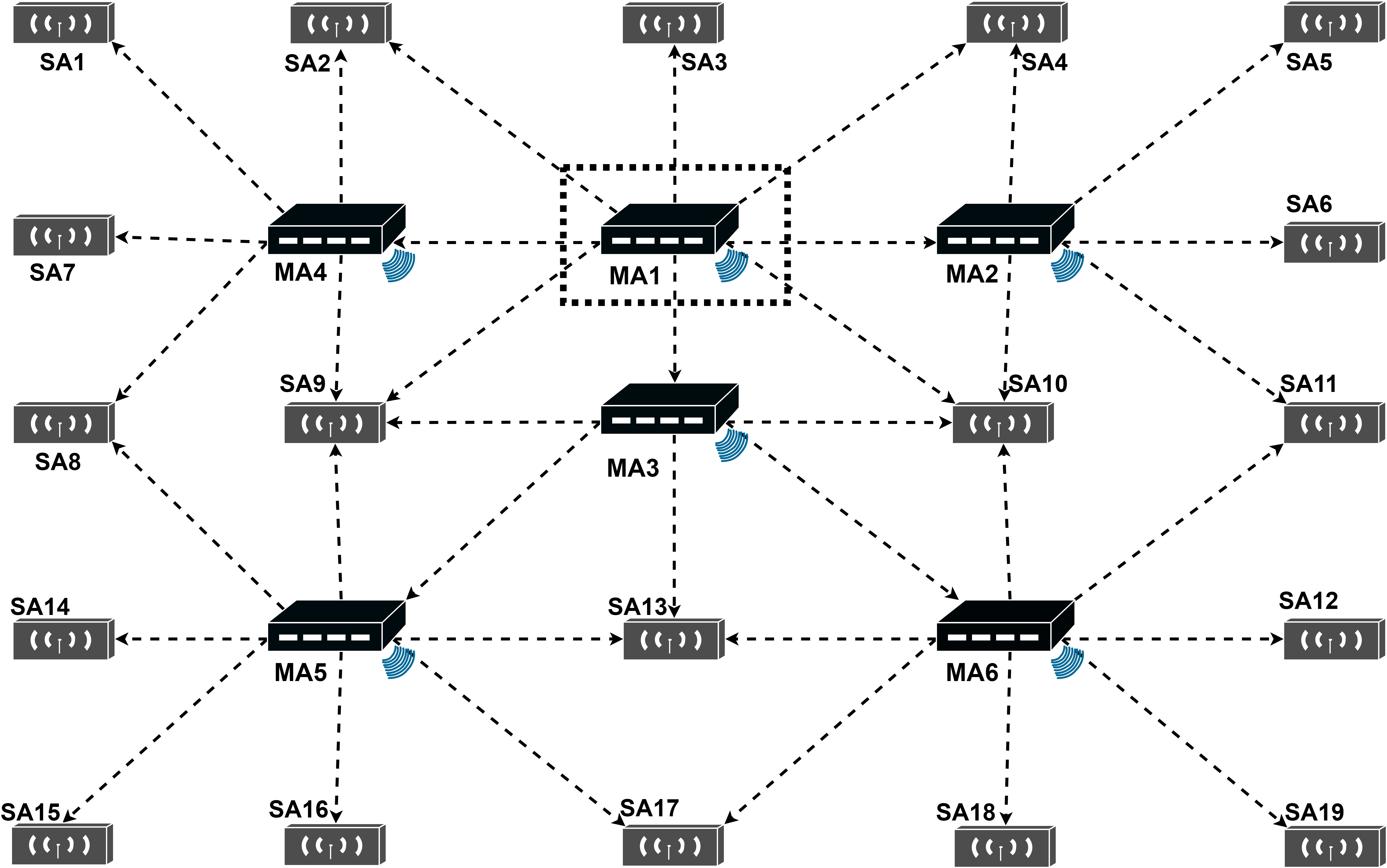}
	\caption{Diagram of WCS with multiple masters.}
	\label{FIG:5}
\end{figure}

In order to cover a large area, it is necessary to employ more than one \textbf{MA}, where each \textbf{MA} is used as a reference to correct the clock drift of its neighboring \textbf{SAs}. We establish a multi-level cascade topological structure of the \textbf{MAs} to coordinate the order of sending \textbf{CCPs} between the primary \textbf{MAs} and the secondary \textbf{MAs} in a complete clock synchronization. 
	
\textbf{MA1} is the primary master anchor. Any \textbf{MA} receiving the \textbf{CCPs} sent by \textbf{MA1} is the secondary master anchor. In addition, the \textbf{MA} receiving the \textbf{CCPs} of the secondary master anchor is level-3 master anchor. So, \textbf{MA2}, \textbf{MA3} and \textbf{MA4} are secondary master anchors, and \textbf{MA5} and \textbf{MA6} are Level-3 master anchors. According to this rule, the cascading model of MA can be obtained as follows: primary master anchor $->$ secondary master anchor  $->$ level-3 master anchor  $->$ level-4 master anchor  $-> ... ->$ level-N master anchor. When a lower level master anchor receiving the \textbf{CCP} from its upper level master anchor, it starts to send the \textbf{CCP} after a short interval (Lag). 
	
Using Figure 5 as an example to illustrate WCS with multiple master anchors. Surrounding the primary master anchor (\textbf{MA1}) are three secondary master anchors (\textbf{MA2}, \textbf{MA3}, \textbf{MA4}) and five slave anchors \textbf{(SA2}, \textbf{SA3}, \textbf{SA4}, \textbf{SA9}, \textbf{SA10}), and they are configured to follow \textbf{MA1}, so their clocks need to be synchronized with \textbf{MA1}. There are five slave anchors (\textbf{SA1}, \textbf{SA2}, \textbf{SA7}, \textbf{SA8}, \textbf{SA9}) around the secondary master anchor \textbf{MA4}, and the slave anchors are configured to follow \textbf{MA4}, so their clocks could be synchronized with \textbf{MA4}. The slave anchors could be configured to follow more than one master anchors, such as \textbf{SA2}, \textbf{SA3}, \textbf{SA4}, \textbf{SA8}, \textbf{SA9}, and \textbf{SA10}. 
	
Based on the topology of multi-level master anchor scheme, We summarize the strategies of WCS with multiple masters: 
\begin{itemize}
	\item Setting every master anchor to send \textbf{CCPs} in a specified interval; 
	\item Choosing one master anchor as the primary master anchor (There is one and only one primary master anchor in a RTLS);
	\item Setting the rest of master anchors as different levels;
	\item Setting the lower level master anchors to follow its upper level master anchor;
	\item Setting the lower level master anchors to send \textbf{CCPs} with different Lags to avoid the collision (for example, \textbf{MA2} delay one Lag, \textbf{MA3} delay two Lags, and \textbf{MA4} delay three Lags);
	\item Setting the slave anchors to follow the master anchor (Ensure the slave anchors could receive the \textbf{CCPs} sent by its master anchor);
	\item Collecting the recorded ToAs of all the anchors and the transmiting timestamps of master anchors;
	\item Using the WCS method shown in Section \textbf{4.1} to synchronize the TDoAs. 
\end{itemize}

\section{Location Estimation Based on TDoA}
\subsection{Model}
TDoA based localization is a common approach used in the UWB system. The positioning system includes several anchors and tags. Assume there is only a single tag to be localized. The tag transmits signals to the anchors periodically. The model of TDoA can be expressed in Figure 6.
\begin{figure}
	\centering
		\includegraphics[scale=.85]{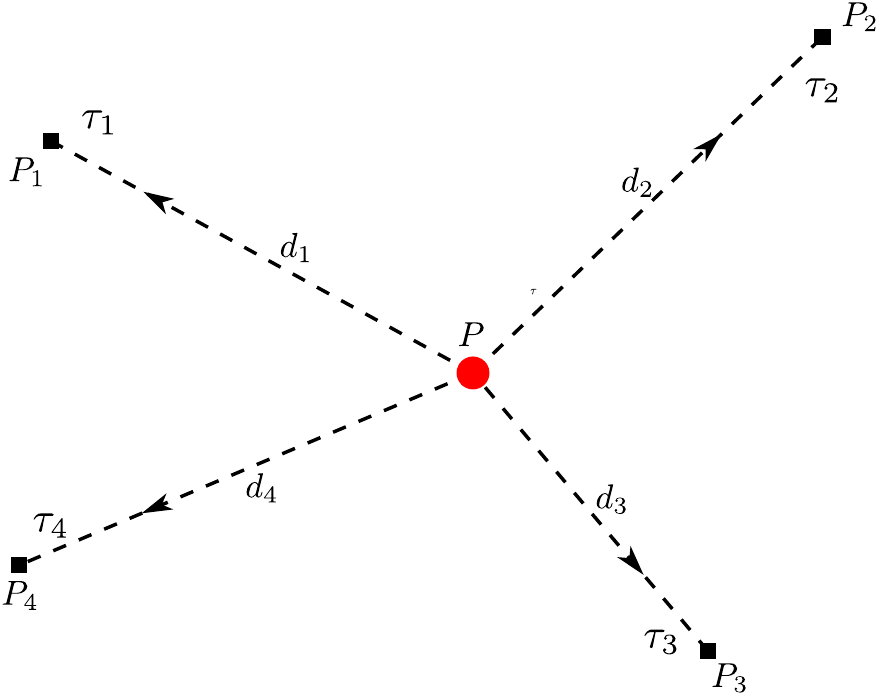}
	\caption{The model of TDoA.}
	\label{FIG:6}
\end{figure}

The tag could transmit the positioning frame to the anchors, and propagate in a straight line to the anchors (Line of sight condition should be satisfied, LoS). For two-dimension positioning, assuming that $p=(x, y)^{T} \in R^{2}$ is the coordinates of the target, where $p_{i}=(x_{i}, y_{i})^{T} \in R^{2}, i=1,2,...,n$ are the coordinates of anchors, and $d_{i}=||p-p_{i} ||$ are the distances between the tag and anchors. 
	
We could get the timestamp \(\tau_{i}\) when the frame is received by the anchors. Assuming that the measurement of \(\hat{\tau}_{i}\) satisfies \(\hat{\tau_{i}}\sim N(\tau_{i}, \delta^{2}_{i})\)\citeneed. The core formula of TDoA is
\begin{equation}
	\hat{d}_{ij} \stackrel{\Delta}{=} \hat{d}_{i}-\hat{d}_{j} = c(\hat{\tau}_{i}-\hat{\tau}_{j}) := c\hat{\tau}_{ij}, \quad \forall i,j = 1,2,\ldots,n,
	\end{equation}
	where \(c\) is the speed of light, the range differences (RD) represent the TDoA measurements, and the distance \(\hat{d}_{ij}\) of here satisfy \(\sum_{i,j} \hat{d}_{ij} \equiv 0\).
	
The aim is to find \(p\), so
\begin{equation}\label{eq:target}
	d_{ij}(p) = \hat{d}_{ij}, \qquad \forall \ i, j=1,2,\ldots,n,
\end{equation}
where \(d_{ij}(p) = d_i(p) - d_j(p) = \norm{p-p_{i}} - \norm{p-p_{j}}\).
	
The inputs of the method are the anchors' coordinates, \(p_{i}\), the measured TDoA, \(\hat{\tau}_{ij}\), and the outputs are the coordinates of the tag, \(p\).
	
Because of the measurement error, the least square condition is usually considered to estimate the position \(\hat{p}\).
\begin{equation}
	\hat{p}=\argmin_p{\sum\limits_{i,j \in 1,\ldots,n}\norm{d_{ij}-\hat{d}_{ij}}}^{2}.
\end{equation}
	
This is a non-convex optimization problem. So the EKF method could be adopted to solve this problem. 

\subsection{The algorithm of EKF}
Location estimation through EKF is available in \cite{khan2013hybrid}, a constant velocity model is selected to describe the RTLS designed in this paper. The state equations and update equations of the EKF model are illustrated in the following formulas. The inputs of EKF include the range difference (RD) $\hat{d_{ij}}$, the coordinates of anchor \(p_{i}\), and the initial position of the Tag \(p\). Meanwhile, the state transfer matrices and other covariance matrices used in the EKF algorithm could be calculated through the references \cite{DBLP:conf/ipin/TiemannEW16a} and \cite{khan2013hybrid}. 

\begin{enumerate}[a.]
\item \textbf{The process model}
$$X_{k}=f\left(X_{k-1}, u_{k-1}\right)+W_{k}$$
$$Z_{k}=h\left(X_{k}\right)+V_{k}$$
	
\item \textbf{Time Update (“Predict”)}
$$\hat{X}_{k}=F_{k}\hat{X}_{k-1}+B_{k}.u_{k}$$
$$P^{\textendash}_{k}=F_{k}P_{k-1}F_{k}^{T}+Q_{k}$$
	
\item \textbf{Measurement Update (“Correct”)}
$$K_{k}=P^{\textendash}_{k} H_{k}^{T}\left(H_{k} P^{\textendash}_{k} H_{k}^{T}+R_{k}\right)^{-1}$$
$$\hat{X}_{k}=\hat{X}^{\textendash}_{k}+K_{k}\left(Z_{k}-h\left(\hat{X}^{\textendash}_{k}\right)\right)$$
$$P_{k}=\left(I-K_{k} H_{k}\right) P^{\textendash}_{k}$$
\end{enumerate}

Where
\begin{itemize}
	\item $X_{k}$ is the true state vector, $X_{k}=[x, y, v_{x}, v_{y}]$, where, $x$ and $y$ represent the coordinates of the tag's position \(p\), $v_{x}$ and $v_{y}$ represent the velocities along the x and y directions;
	\item $Z_{k}$ is the observation vector, and $h\left(X_{k}\right)$ is the observation vector which can be defined as:
	\[h\left(X_{k}\right)=\begin{bmatrix} \hat{d_{1j}} \\ \hat{d_{2j}} \\ ... \\ \hat{d_{ij}} \end{bmatrix},{i \in 2,\ldots,n}, j=1;\]
	\item $\hat{X}_{k}$ a priori state vector of a posteriori estimated state vector $\hat{X}_{k-1}$, where $\hat{d_{ij}}$ can be calculated by formula (5);
	\item $W_{k}$ represents the process noise, and $V_{k}$ represents the observation noise vector;
	\item $B_{k}$ is the input matrix, and $u_{k}$ is the input to the system;
	\item $P_{k}$ is the estimated covariance matrix, and $K_{k}$ is the Kalman gain martix;
	\item $Q_{k}$ represents the covariance matrix, and $R_{k}$ represents the covariance matrix related to the observation vector;
	\item $F_{k}$ represents the linearized state transition matrix, which can be defined as follows: 
	\[F_{k}=\begin{bmatrix} 1 & 0 & 0.01 & 0 \\ 0 & 1 & 0 & 0.01 \\ 0 & 0 & 1 & 0 \\ 0 & 0 & 0 & 1 \end{bmatrix};\]
	\item $H_{k}$ represents the Jacobian matrix related to expected measurements, which can be defined as follows:
	\[H_{k}=\begin{bmatrix} \frac{\hat{x}-x_{1}}{ED({\hat{p},p_{1}})} & \frac{\hat{y}-y_{1}}{ED({\hat{p},p_{1}})} & 0 & 0 \\ ... & ... & ... & ... \\ \frac{\hat{x}-x_{i}}{ED({\hat{p},p_{i}})} & \frac{\hat{y}-y_{i}}{ED({\hat{p},p_{i}})} & 0 & 0 \end{bmatrix},\] where $h(\hat{X}^{\textendash}_{k})=ED({\hat{p},p_{i}})$ is the estimated Euclidean distance between the tag and the i-th UWB anchor, and $\hat{p}$ is the estimate position which is defined at formula (7).
	$$ED({\hat{p},p_{i}})=\sqrt{(\hat{x}-x_{i})^{2}+(\hat{y}-y_{i})^{2}}.$$
\end{itemize}

The details of the location estimation method based on EKF is demonstrated in \textbf{Algorithm 1}.
\begin{algorithm}[htb]
	\caption{Solution of EKF}
	\label{alg:example}
	\begin{algorithmic}[1] % show the line number
		\REQUIRE ~~\\ % algorithm input parameters：Input
		Coordinate $p=(x_{i},y_{i})$ for each anchor and measured distance difference $\hat{d}_{ij}$ between tag and each pair of anchors;\\
		\ENSURE ~~\\ % algorithm output：Output
		The expected position $\hat(p)=(x,y)$ of the tag;
		\STATE {Initialize the state vector $X_{k}$, $F_{k}$, $P_{k}$, $Q_{k}$, $R_{k}$, $H_{k}$}
		\STATE \quad {$\hat{X}_{k}\leftarrow F_{k}\hat{X}_{k-1}, B_{k}.u_{k}$}
		\STATE \quad {$P^{\textendash}_{k}\leftarrow F_{k}P_{k-1}F_{k}^{T}+Q_{k}$}
		\STATE \quad {$K_{k}\leftarrow P^{\textendash}_{k} H_{k}^{T}\left(H_{k} P^{\textendash}_{k} H_{k}^{T}+R_{k}\right)^{-1}$}
		\STATE \quad {$\hat{x}_{k}\leftarrow \hat{X}^{\textendash}_{k}+K_{k}\left(Z_{k}-h\left(\hat{X}^{\textendash}_{k}\right)\right)$}
		\STATE \quad {$P_{k}\leftarrow \left(I-K_{k} H_{k}\right) P^{\textendash}_{k}$}
	\end{algorithmic}
\end{algorithm}

\section{Time-base Selection Strategy and Anchor Deployment Scheme}
\subsection{Time-base Selection Strategy}
We need time-base selection strategy to cover different anchor deployment schemes in the TDoA-based real-time localization system, where time-base means the reference anchor's ToA which will be selected to get the TDoA ($TDoA = ToA_{i} - ToA_{reference}$). The anchor deployment based on WCS with a single master is illustrated in Figure 7, we choose the recorded timestamp of the single master anchor, \textbf{MA1}, as the time-base of TDoA. 
	
We define the time-base selection rules with multiple master anchors as follows: 
\begin{itemize}
	\item If the \textbf{CCP} received by a \textbf{SA} came from a same \textbf{MA}, the recorded timestamp of the \textbf{MA} is selected as the time-base of TDoA. 
	\item If the \textbf{CCP} received by a \textbf{SA} came from two or more \textbf{MAs}, then the recorded timestamp of the \textbf{SA} is selected as the time-base of TDoA. 
\end{itemize}

As an example, the anchor deployment based on WCS with multiple masters is demonstrated in Figure 8, when the Tag is located in Cell-1, \textbf{SA2} could receive the \textbf{CCPs} from both \textbf{MA1} and \textbf{MA3}, so the recorded timestamp of \textbf{SA2} is selected as the time-base of TDoA; when the Tag moves to Cell-2, the \textbf{SAs} in this region can only receive \textbf{CCPs} from \textbf{MA2}, so the recorded timestamp of \textbf{MA2} is selected as the time-base of TDoA. 
	
\subsection{Anchor Deployment Scheme}
In a real deployment scenario, the necessary deployment rules need to be followed to ensure positioning accuracy. Corresponding to the two WCS schemes proposed in Section \textbf{4}, we present two typical reference deployment schemes in practical scenarios. 

The anchor deployment based on WCS with a single master is illustrated in Figure 7. We consider the case of a single master anchor covering a larger area, so the number of anchors go up to eight. Besides, to ensure smooth communication between the master anchor and all the surrounding slave anchors, we put the master anchor at the center of the positioning area. 
\begin{figure}
	\centering
		\includegraphics[scale=.08]{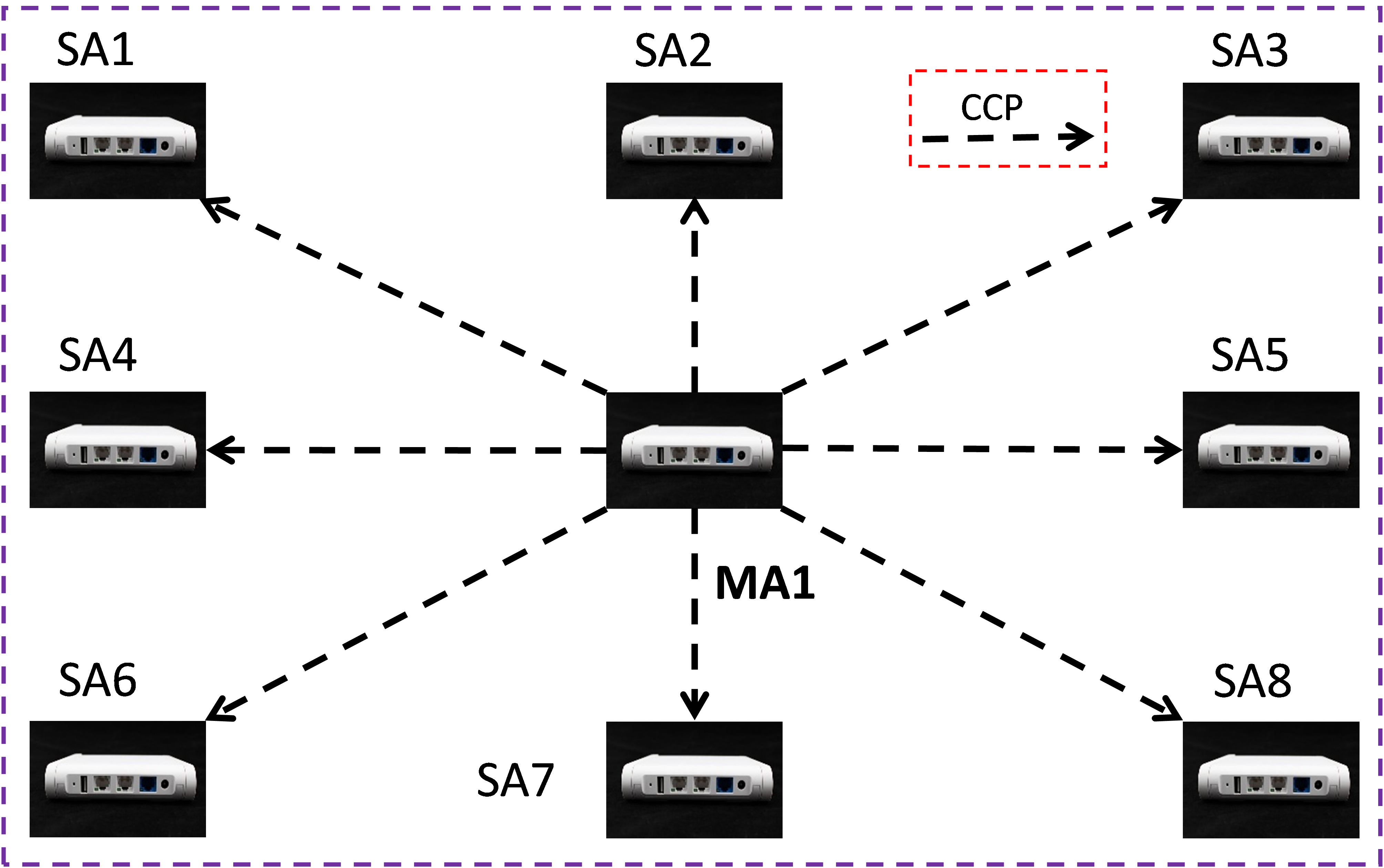}
	\caption{Diagram of anchor deployment with a single master.}
	\label{FIG:7}
\end{figure}

In Figure 8, we discuss the situation of the location area with multiple master anchors. To make it easy to describe, we assume that the tracking space is divided into four separate areas, including one primary master anchor MA1, four secondary master anchors, MA2-MA5, eleven slave anchors, SA1-SA11. The dotted line indicates the CCP, representing the synchronous message. In the case of the current reference anchor deployment, the WCS with multiple masters scheme is adopted.
\begin{figure}
	\centering
		\includegraphics[scale=.062]{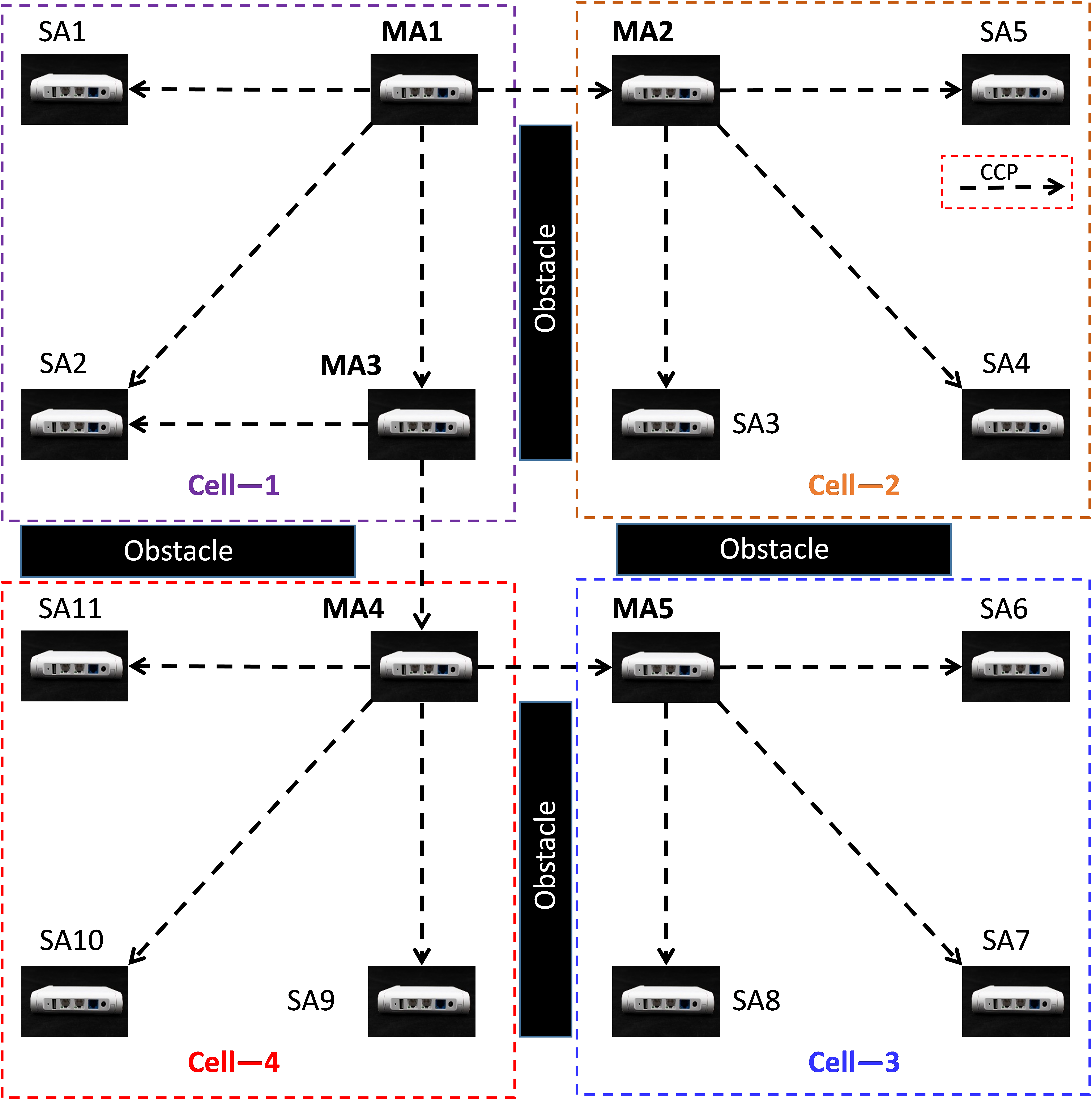}
	\caption{Diagram of anchor deployment with multiple masters.}
	\label{FIG:8}
\end{figure}

\subsection{Common Deployment Rules:}
The following seven rules extracted from our experiments should be considered when deploying the anchors.
\begin{enumerate}[a.]
	\item Keep a LoS (line of sight) between a master anchor and slave anchors (at least 3 slave anchors).
	\item The anchors should be installed above the localized objects. Assure a clear line of sight between tags and anchors. Do not hide the Tag behind materials that attenuate the radio signal like water, a human body, or metal parts.
	\item Mount anchors (surrounding Master anchor) ideally at the same height (keep a variation of 1 meter maximum).
	\item Keep anchors detached away from walls or ceilings (ideally 50 cm but not less than 15 cm – shorter detachment may cause higher signal attenuation and inaccuracys due to reflections).
	\item Keep a square geometry when designing anchor deployment. The minimum distance between anchors should be longer than 3 meters. The location area should be bigger than 3m x 3m.
	\item Orient the anchors such that their radiation capabilities are satisfactory.
\end{enumerate}

\subsection{Dilution of Precision (DoP) Guided Deployment}
The DoP model could be used for evaluating the relationship between the anchors' placement geometry with the positioning accuracy of the RTLS. The DoP model could be adapted to measure various positioning systems' performance and is independent of communication technologies and modes. We have studied the DoP for the TDoA technique with respect to anchor deployment in \citep{Zhang2021DilutionOP}. 

The DoP provides a gain factor that is numerically dimensionless and represents the relationship between the measurement error at a given position and the geometry of the anchors. It should be noted that the relationship between anchor spacing and DoP is fragile, but their geometric structure will have a specific influence. So it's worth discussing how to deploy anchors based on the DoP. We can adopt the most practical horizontal DoP (HDoP) because 2-D positioning is used much more frequently than 3-D positioning in most scenarios. This work mainly take square anchor geometry as suggested by. 

\section{Experiments and Performance Analysis}
\subsection{Introduction of the Experimental Environment}
According to the UWB-based Wireless Positioning Network (U-WPN) proposed in Section \textbf{3}, a test experiment is set up to analyze the proposed approach's positioning accuracy and stability, which is shown in Figure 8. We deploy four anchors in a conference room. They are fixed on the four vertices of a rectangle with 6m's length and 4m's width. The RTLS we have built includes four anchors, two tags, one router, one PoE Switch, and a \textbf{CLE}. The test process is as follows: Four anchors are fixed on the tripods with the same height of 1.80m. The sending periods of \textbf{CCP} and \textbf{Blink} are set respectively.
\begin{itemize}
	\item A CCP is transmitted from the master anchor every 150 milliseconds.
	\item A blink is transmitted from the Tag every 100 milliseconds.
\end{itemize}
\begin{figure}
	\centering
		\includegraphics[scale=.05]{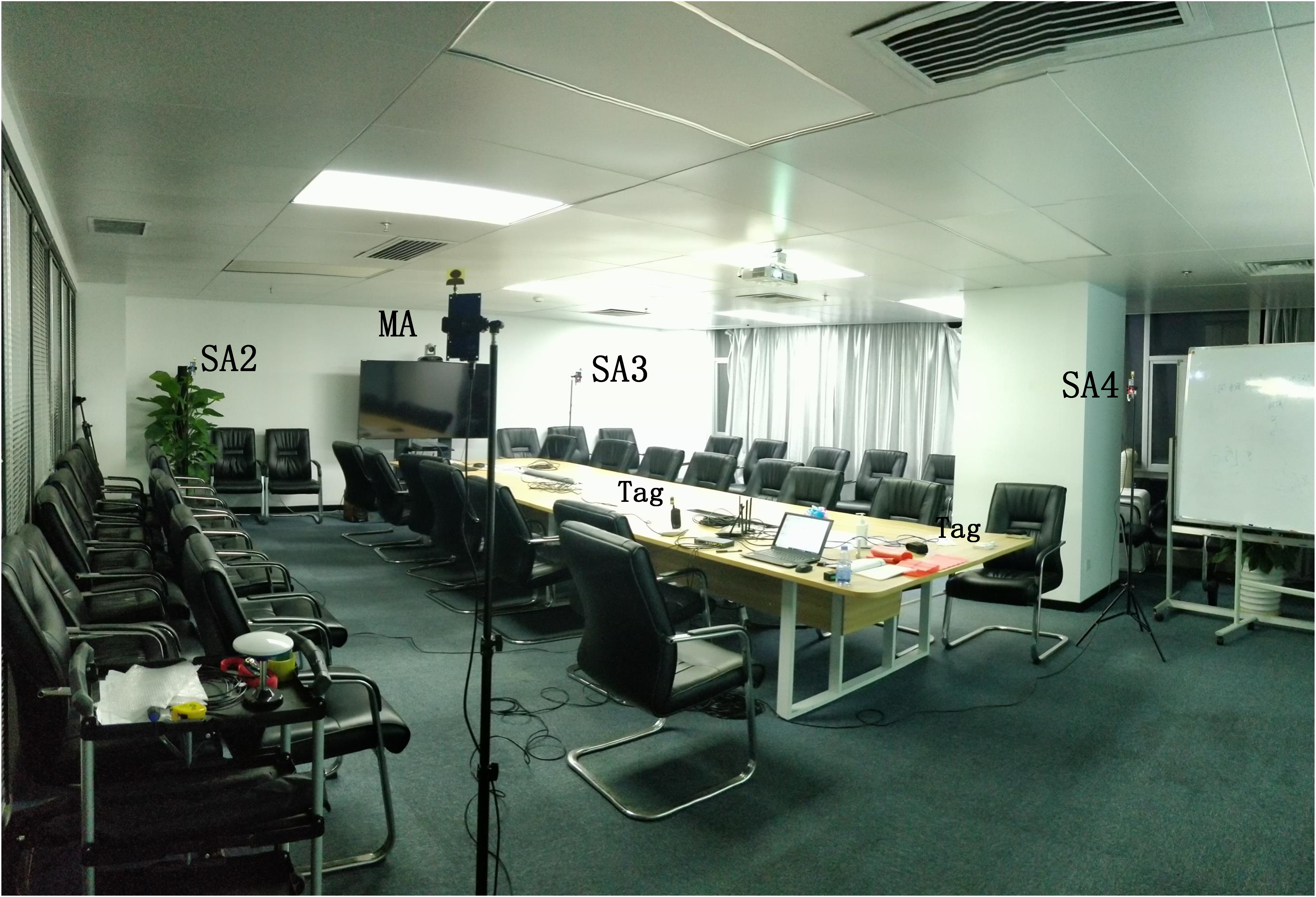}
	\caption{System implementation.}
	\label{FIG:9}
\end{figure}

\subsection{Performance of WCS Scheme with a Single Master Anchor}
The results of the WCS scheme with a single master anchor are shown in Figures 10 and 11. \textbf{SA2-MA}, \textbf{SA3-MA}, and \textbf{SA4-MA} represent the TDoA between the master anchor and three slave anchors respectively. It is obvious that, the original synchronized TDoA data demonstrated in Figure 10, are with large fluctuations and several nanoseconds synchronization errors. Such TDoA data cannot be directly used for precise positioning. We used the Kalman Filter (KF) to process the original synchronized TDoA data, and the results were illustrated in Figure 11. It is obvious that KF greatly reduces the influence of noise on the WCS scheme. 
\begin{figure}
	\centering
		\includegraphics[scale=.4]{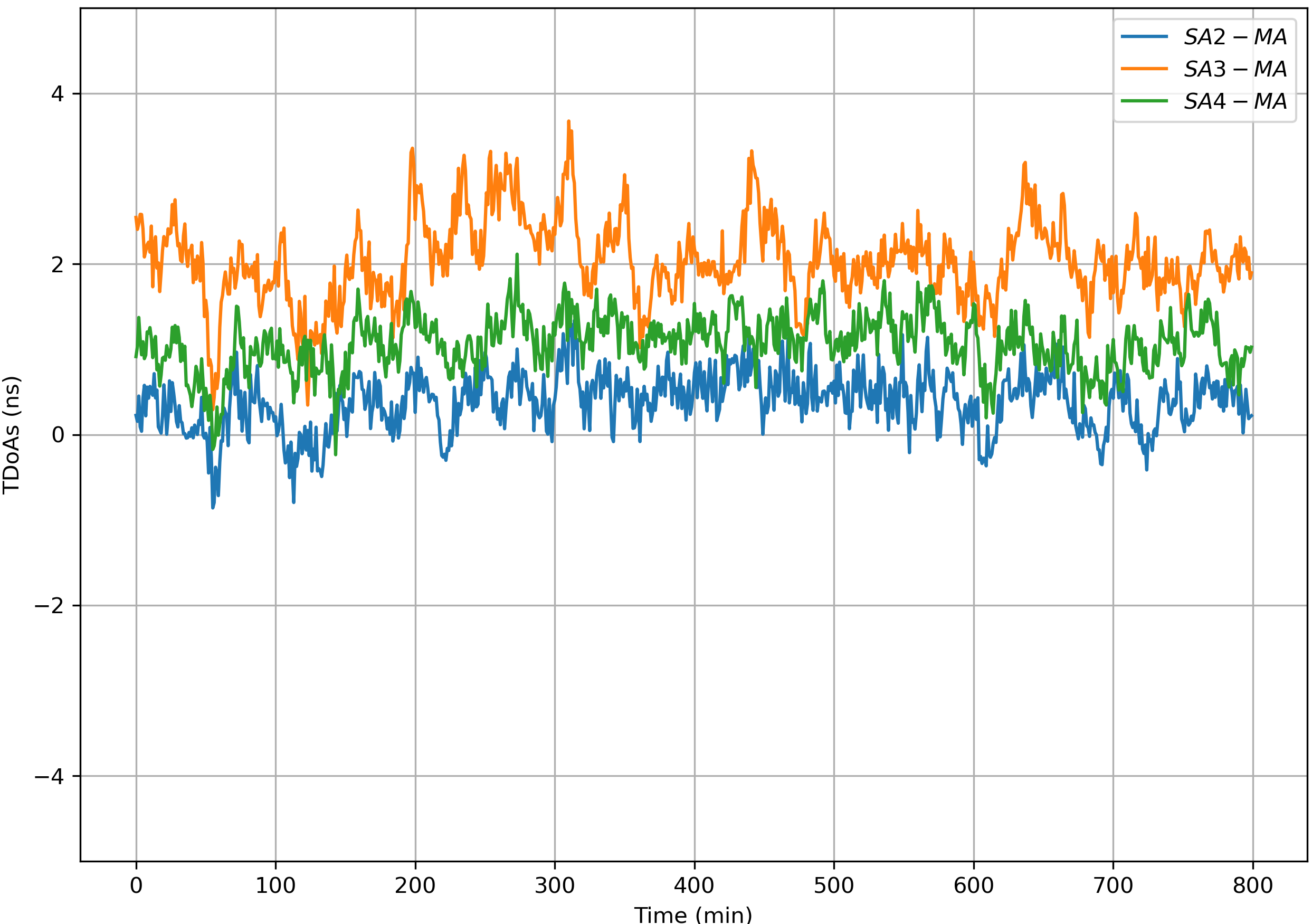}
	\caption{Performance of WCS with a signal master anchor.}
	\label{FIG:10}
\end{figure}

\begin{figure}
	\centering
		\includegraphics[scale=.4]{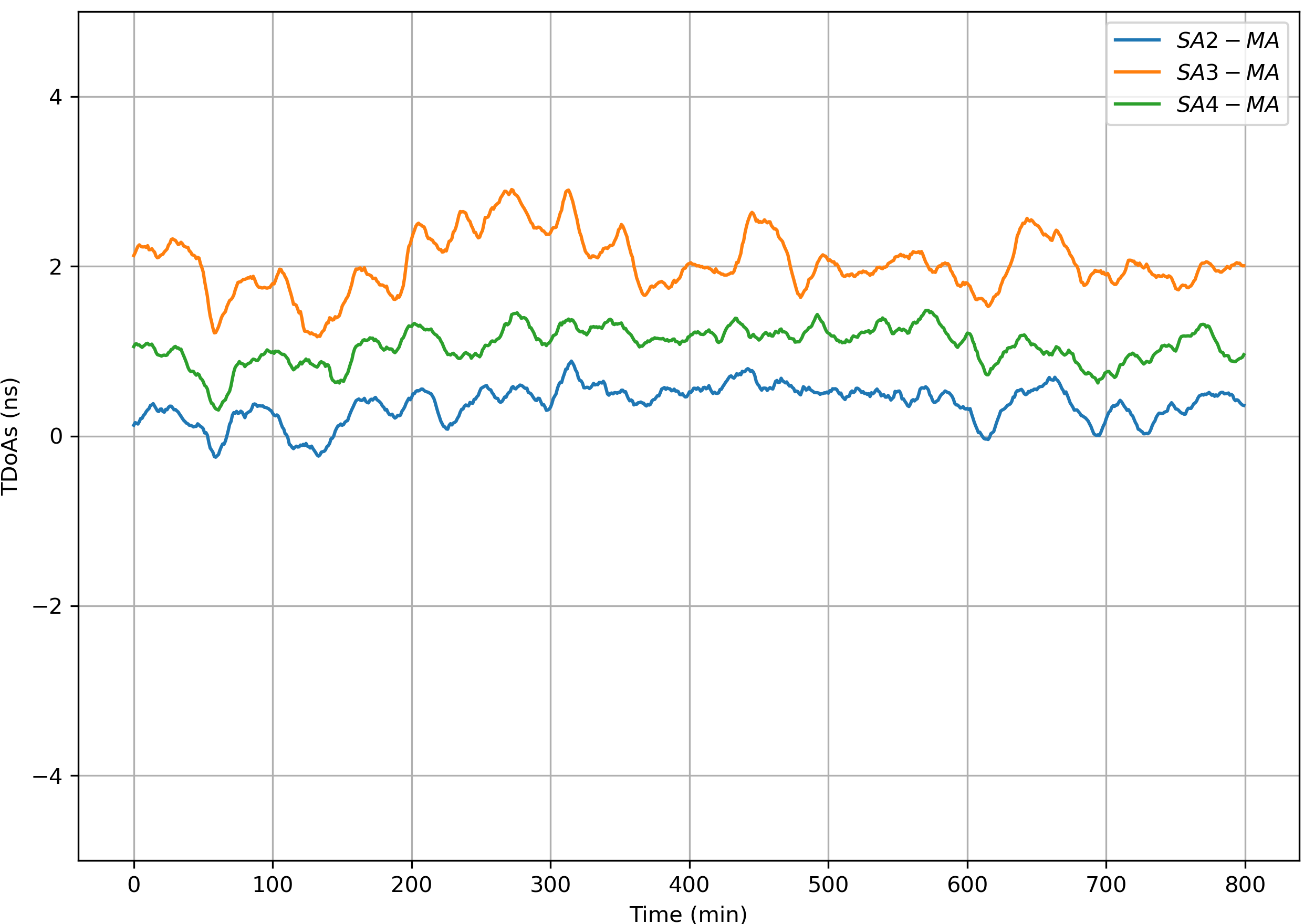}
	\caption{Performance of WCS with a signal master anchor based on KF.}
	\label{FIG:11}
\end{figure}

For each \textbf{SA}, the Blink RX timestamps are corrected with the wireless clock synchronization algorithm. The corrected timestamps, which follow its corresponding master recorded timestamp, are used for calculating the TDoA of each blink message between the \textbf{MA} and each \textbf{SA}. The standard deviations of the TDOA for the deployed three slave anchors in the test environment were 0.18 ns, 0.19 ns, and 0.14 ns throughout an 800-minute test. The average value of these deviations is not more than 200 ps, representing a standard deviation of position less than 6cm.

\subsection{Performance of WCS Scheme with Multiple Master Anchors}
The results of the WCS scheme with multiple master anchors are shown in Figures 12 and 13. \textbf{SA2-MA1}, \textbf{SA3-MA1}, \textbf{MA6-MA1}, \textbf{SA4-MA6}, and \textbf{SA5-MA6} represent the TDoAs between the master anchors and slave anchors respectively.The original synchronized TDoA data demonstrated in Figure 12, and the results after applied KF were illustrated in Figure 13. 
\begin{figure}
	\centering
		\includegraphics[scale=.4]{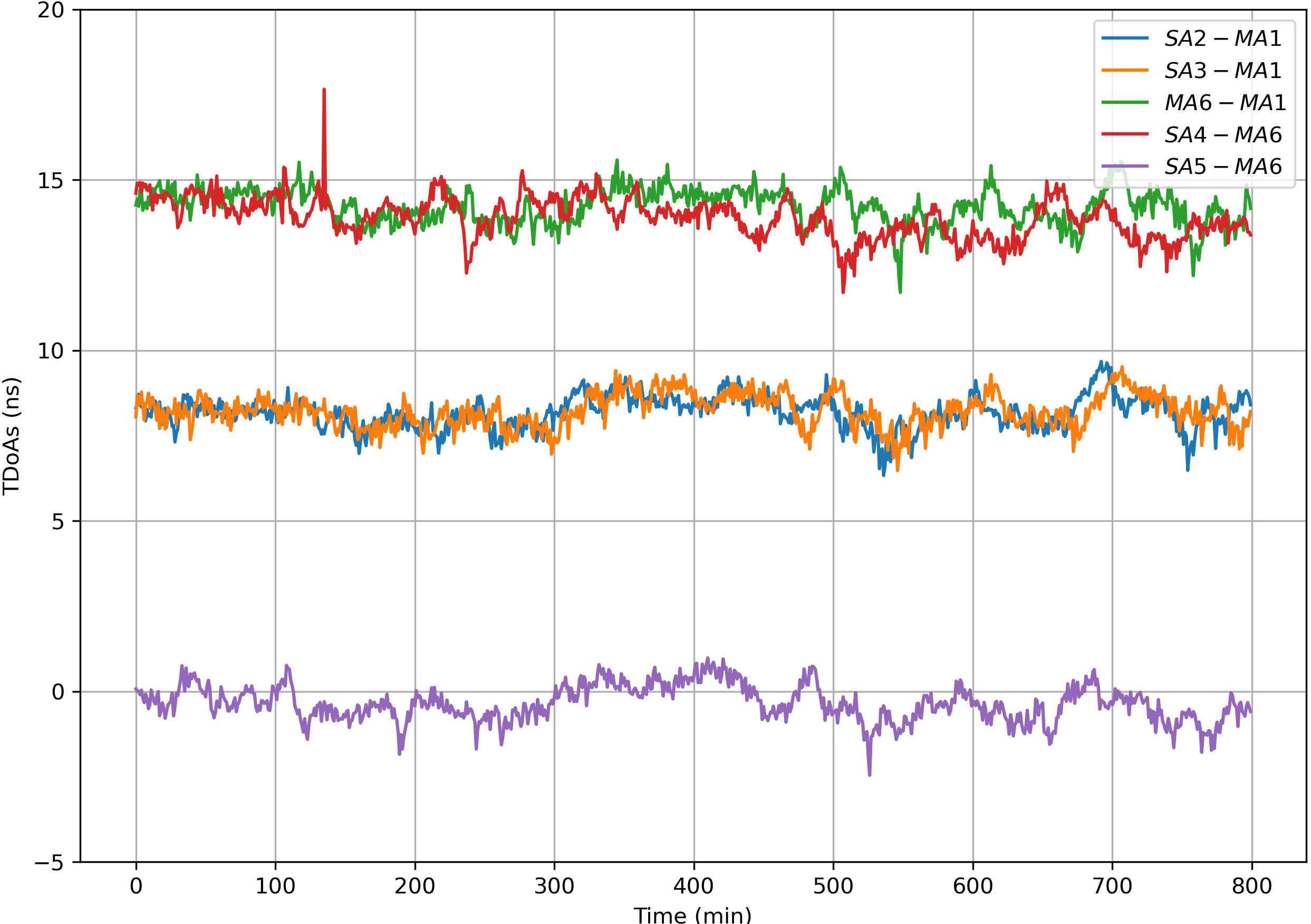}
	\caption{Performance of WCS with multiple master anchors.}
	\label{FIG:12}
\end{figure}

\begin{figure}
	\centering
		\includegraphics[scale=.4]{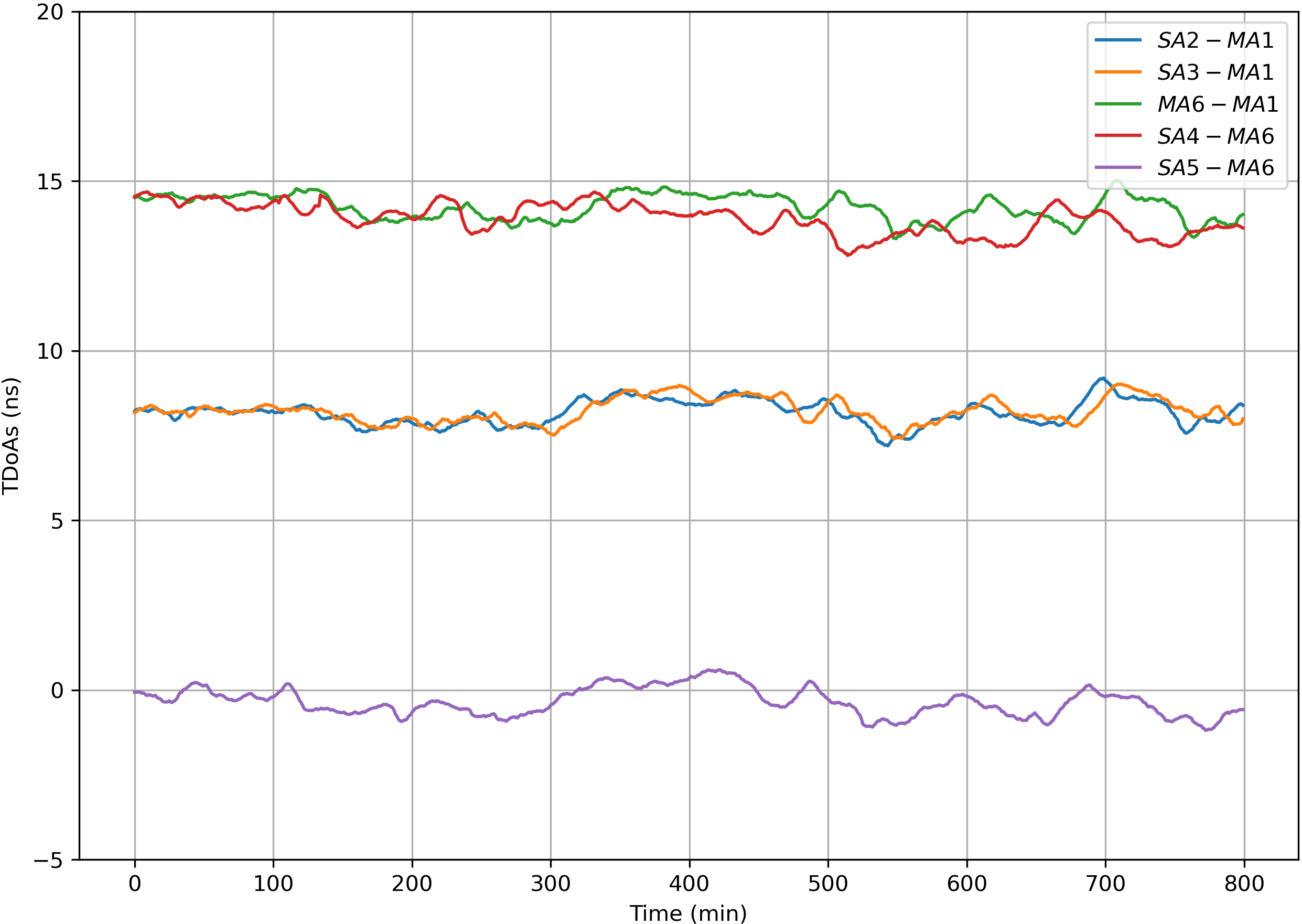}
	\caption{Performance of WCS with multiple master anchors based on KF.}
	\label{FIG:13}
\end{figure}

\subsection{Performance of RTLS with a Single Master Anchor}
We use an extended Kalman filter (EKF) in the RTLS to estimate the tags' positions, and the test results are demonstrated in Figure 14.
\begin{figure}
	\centering
		\includegraphics[scale=.05]{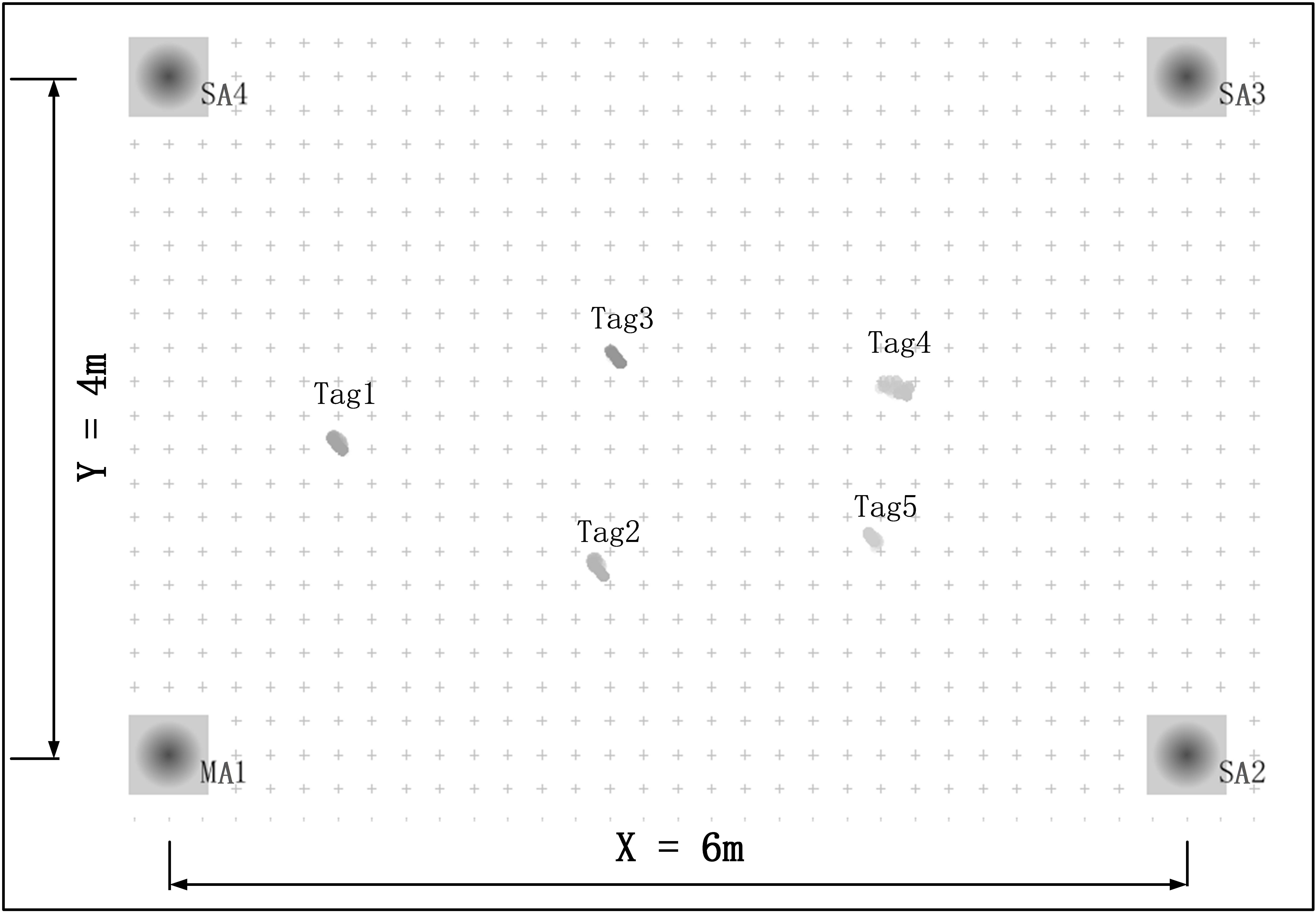}
	\caption{Result of RTLS with a single master anchor.}
	\label{FIG:14}
\end{figure}
The test results of tag tracking with a single master anchor are illustrated in Figure 15. 
\begin{figure}
	\centering
		\includegraphics[scale=.055]{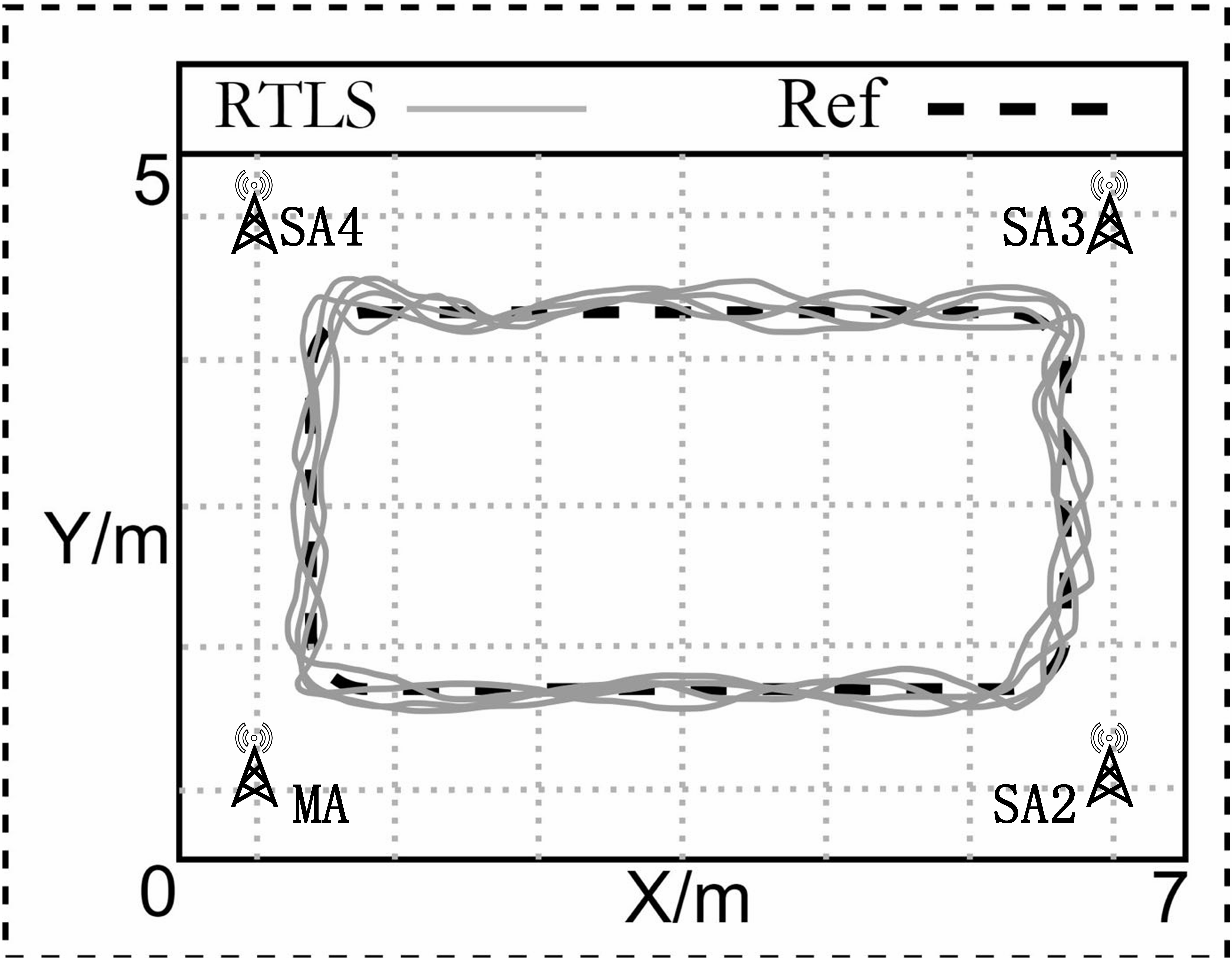}
	\caption{Result of tag tracking with a single master anchor.}
	\label{FIG:15}
\end{figure}

\subsection{Performance of RTLS with Multiple Master Anchors}
Another test experiment is set up to analyze the proposed approach's positioning accuracy and stability, which is shown in Figure 16. The test is conducted in a 9.88-meter wide and 11.02-meter long Hall. We deploy two master anchors (\textbf{MA1}, and\textbf{ MA6}) and four slave anchors (\textbf{SA2}, \textbf{SA3}, \textbf{SA4}, and \textbf{SA5}). Also there are 5 tags (\textbf{Tag1}, \textbf{Tag2},..., \textbf{Tag5}) deployed in this area. 
\begin{figure}
	\centering
		\includegraphics[scale=.055]{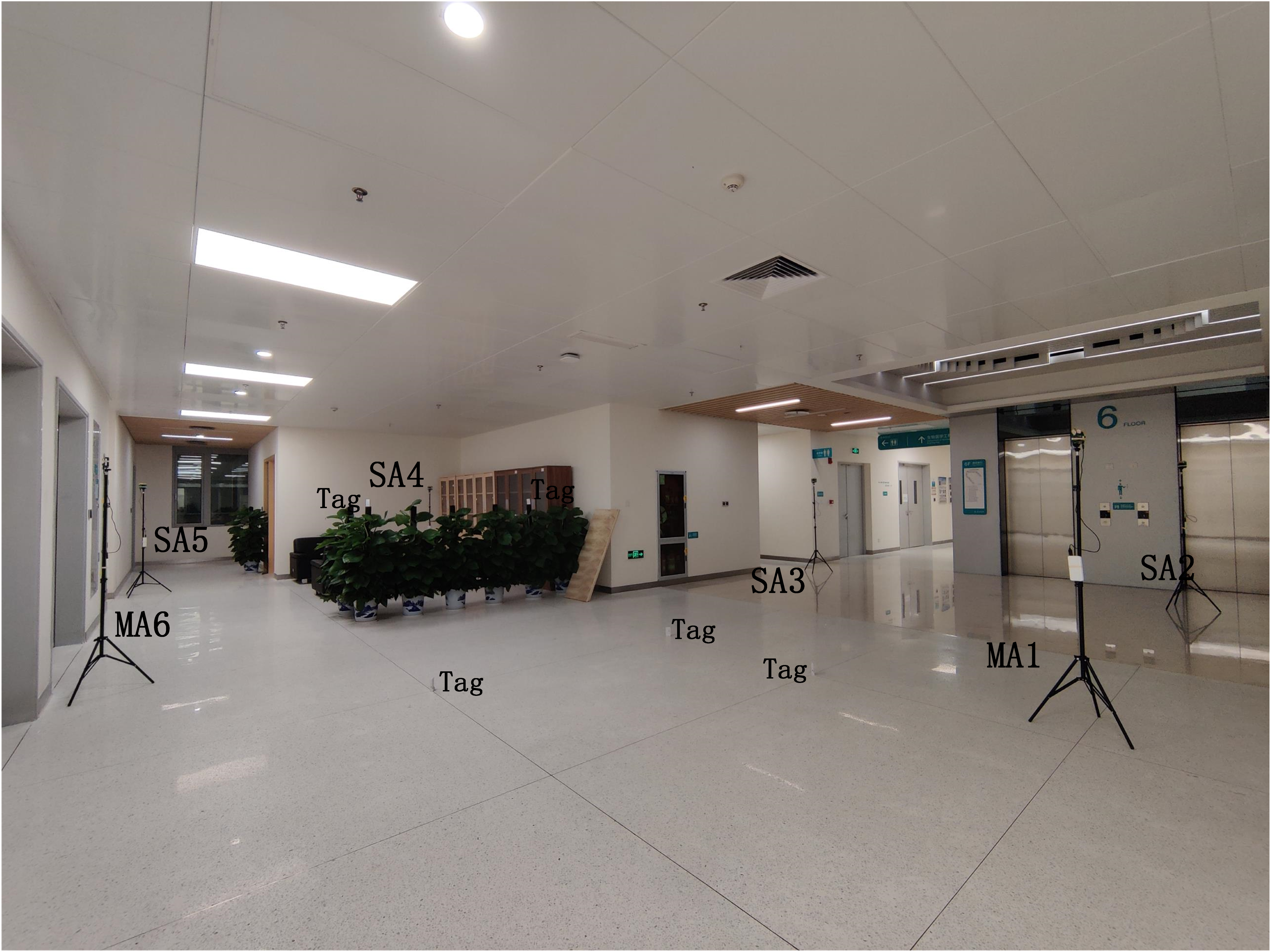}
	\caption{System implementation with multiple master anchors.}
	\label{FIG:16}
\end{figure}

We use the RTLS built in this paper to estimate the tags' positions in a static scenario, and the test results are demonstrated in Figure 17. 
\begin{figure}
	\centering
		\includegraphics[scale=.055]{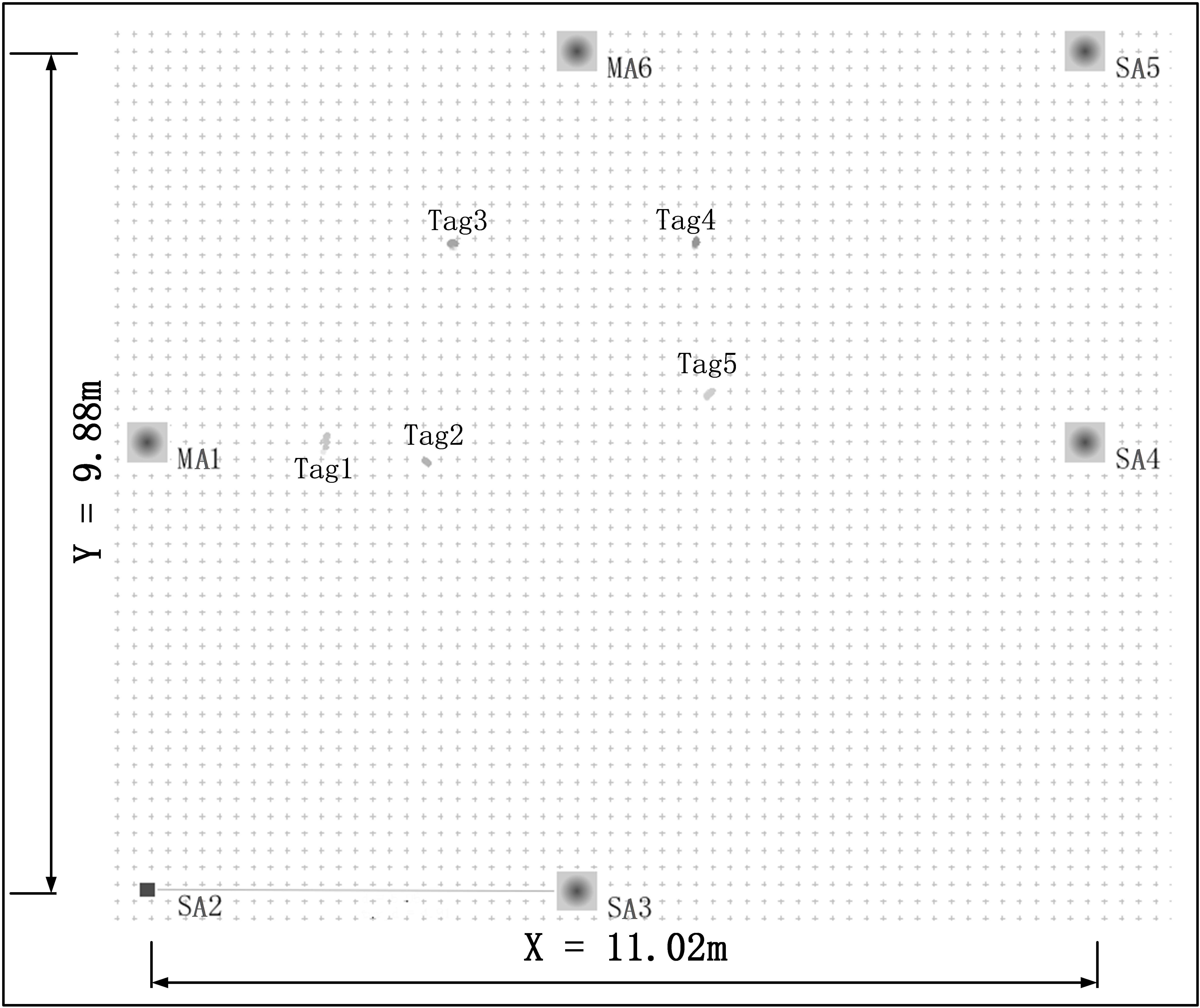}
	\caption{Result of RTLS with multiple master anchors.}
	\label{FIG:17}
\end{figure}

The test results of tag tracking with multiple master anchors are demonstrated in Figure 18. 
\begin{figure}
	\centering
		\includegraphics[scale=.05]{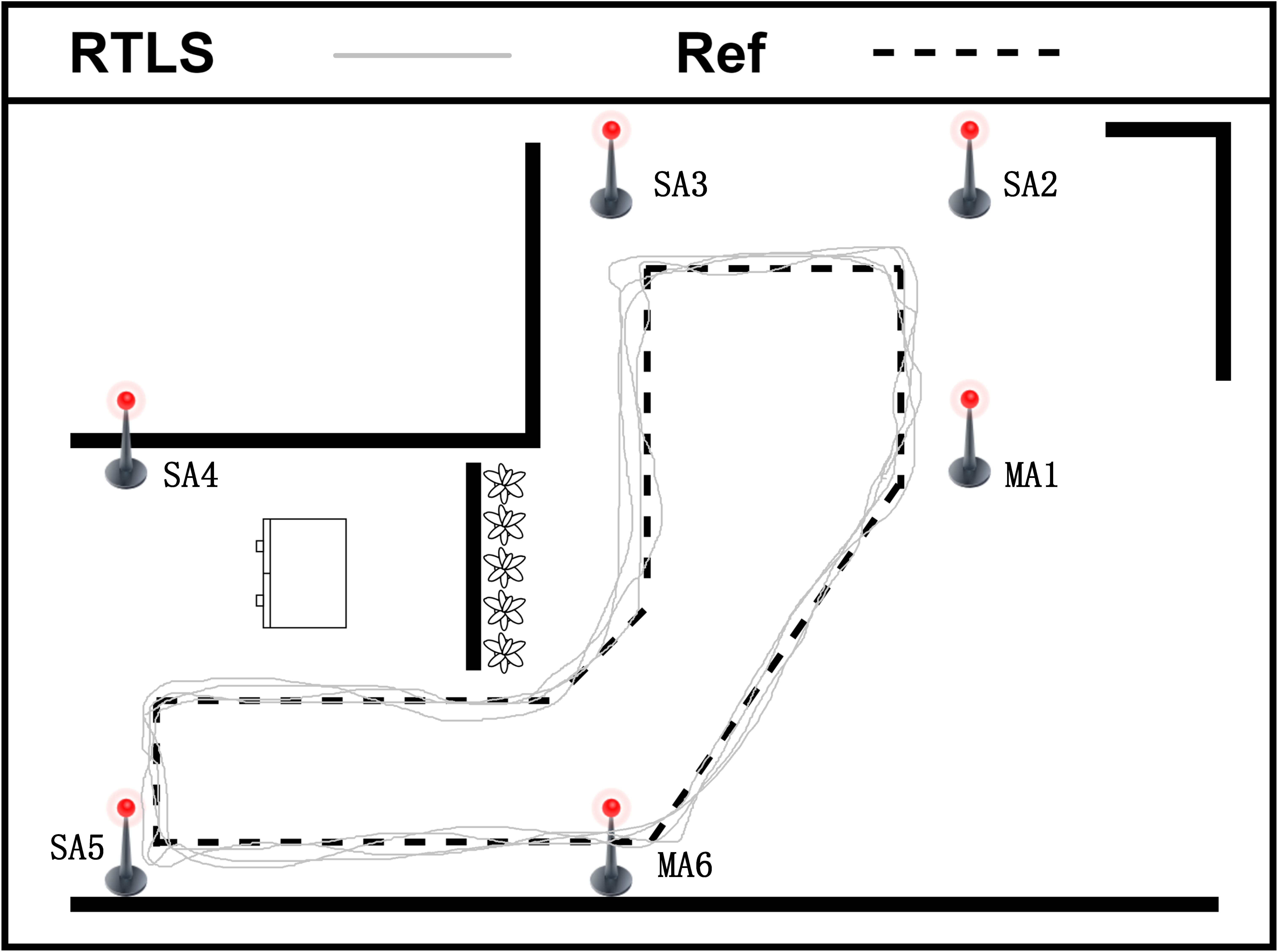}
	\caption{Result of tag tracking with multiple master anchors.}
	\label{FIG:18}
\end{figure}

To illustrate the effectiveness of the adopted EKF scheme during static situation and tag tracking, we conduct the above experiments. The experimental results as shown in Figure 14, Figure 15, Figure 17, and Figure 18, illustrate that the RTLS designed in this paper has stable performance in tracking and has high positioning accuracy while ensuring positioning continuity. We can achieve less than 10cm the positioning error when the tag is fixed and less than 30cm when the tag is moving. Limited by the hardware's clock resolution, it is hard to completely eliminate clock bias and positioning errors.

\subsection{Comparison of Positioning Performance of Different Schemes}
Table 1 demonstrates the positioning performance of eleven schemes using TDoA method. ATLAS's sheme amd Bitcraze's scheme are open source, and the other eight schemes are commercial. According to the data provided by each solution provider, the static positioning accuracy is about 10cm, and the dynamic positioning accuracy is about 30cm. Compared with our proposed scheme, the positioning accuracy is almost the same, because any mature solution uses unique algorithms to eliminate external errors, so the positioning accuracy depends on the performance of the UWB RF chip. The performance of our proposed scheme is consistent with the mainstream schemes, and it has flexible expansion in the aspect of multi-anchor cascade. 
\begin{table}
	\caption{Positioning Performance of Different Schemes}	
	\label{table_time}
	\begin{tabular*}{\tblwidth}{@{} LLLL@{} } 
		\toprule   
		Scheme & Source &Static &  Dynamic \\  
		\midrule   
		SUSTechRTLS &  Lab Research & 5- 10cm &  about 30cm \\  
		DecaWaveRTLS &  Commercial &  about 10cm &  about 30cm \\ JINGWEI & Commercial &  about 10cm & 10- 30cm \\
		ATLAS & Open Source &  about 10cm & 20-30cm \\
		woxuwireless & Commercial &  about 10cm &  about 30cm \\
		Bitcraze & Open Source &  about 10cm &  10-30cm \\
		EHIGH & Commercial &  about 10cm &  about 30cm \\
		Sewio & Commercial &  less 10cm &  about 30cm \\
		Localsense & Commercial &  about 10cm &  10-30cm \\
		zebra & Commercial &  about 10cm &  10-30cm \\
		ubitraq & Commercial &  about 10cm &  10-30cm \\
		\bottomrule  
	\end{tabular*}	
\end{table}

\section{Conclusion and Future Work}
This paper reviews the existing technologies and solutions for precise positioning, analyzes the advantages of UWB technology, discusses the current available solutions of mainstream manufacturers in UWB. Then we summarizes the algorithms and solutions needed to build a real-time localization system based on UWB. To address the deficiencies and challenges of the existing solutions, a comprehensive UWB-based RTLS is proposed. 
	
First of all, we design and implement the hardware and software of anchors and tags based on UWB, which have good performance and low power consumption. Then, we propose the new wireless clock synchronization (WCS) method, and define the time-base selection strategy for TDoA algorithm in signal master anchor and multiple master anchors systems. Meanwhile, the EKF method for solving TDoA issues is introduced for nonlinear dynamic systems, and it is useful in moving target tracking with the real-time positioning accuracy up to 30cm. Finally, we discuss the relationship between anchor deployment and positioning accuracy. 
	
At present, DW1000 chip follows the standard of  IEEE 802.15.4-2011. Meanwhile, the IEEE 802.15.4z standard is announced to support DW3000, Apple U1, and NXP SR100T. The IEEE 802.15.4z defines new features based on the original standard, with enhanced security, lower power consumption, and longer transmission distance. We will continue our research based on the new standard, including TDoA, arrival of angle (AoA), and phase difference of arrival (PDoA). 

\printcredits

\section*{Acknowledgment}
This research is supported in part by the National Natural Science Foundation of China (Grant No. 92067109, 61873119); and in part by the Scientific Research Foundation of Science and Technology on Near-Surface Detection Laboratory of China (Grant No. TCGZ2018A006); and in part by the Science and Technology Planning Project of Guangdong Province(Grant No. 2021A0505030001); and in part by the Educational Commission of Guangdong Province (Grant No. 2019KZDZX1018).


\bibliographystyle{cas-model2-names}
\bibliography{jnca-rtls}

\bio[width=10mm,pos=l]{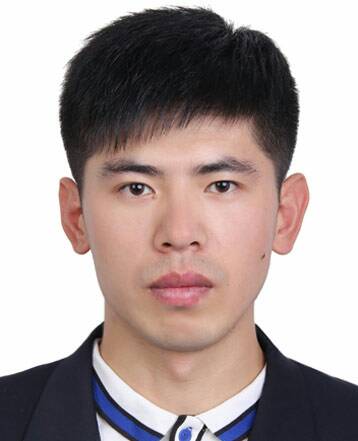}
{Fengyun Zhang} is a PhD student in the Department of Computer Science and Engineering at Southern University of Science and Technology (SUSTech). His research interests include UWB indoor location and industrial control network protocol reverse engineering. He received his B.E. degree in 2014 with the major of electronic and information engineering, and his M.E. degree in 2017 with the major of signal and information processing both from Southwest University, China respectively. 
\endbio

\bio[width=10mm,pos=l]{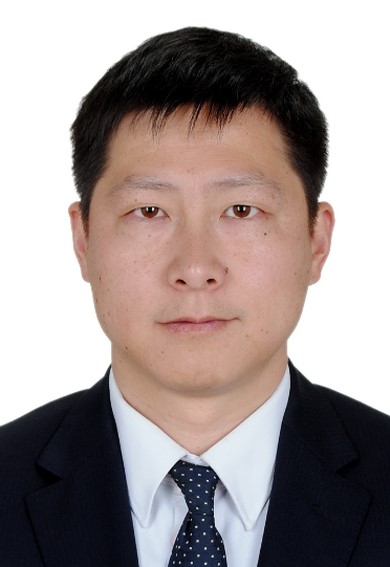}
{Li Yang} received the Ph.D. degree in mechanical and electronic engineering from The Nanjing University of Science and Technology, China, in 2013. He is currently an associate professor of the Army Engineering University of PLA. His main interests are Ultra-Wideband (UWB) technology and Ad hoc network.
\endbio

\bio[width=10mm,pos=l]{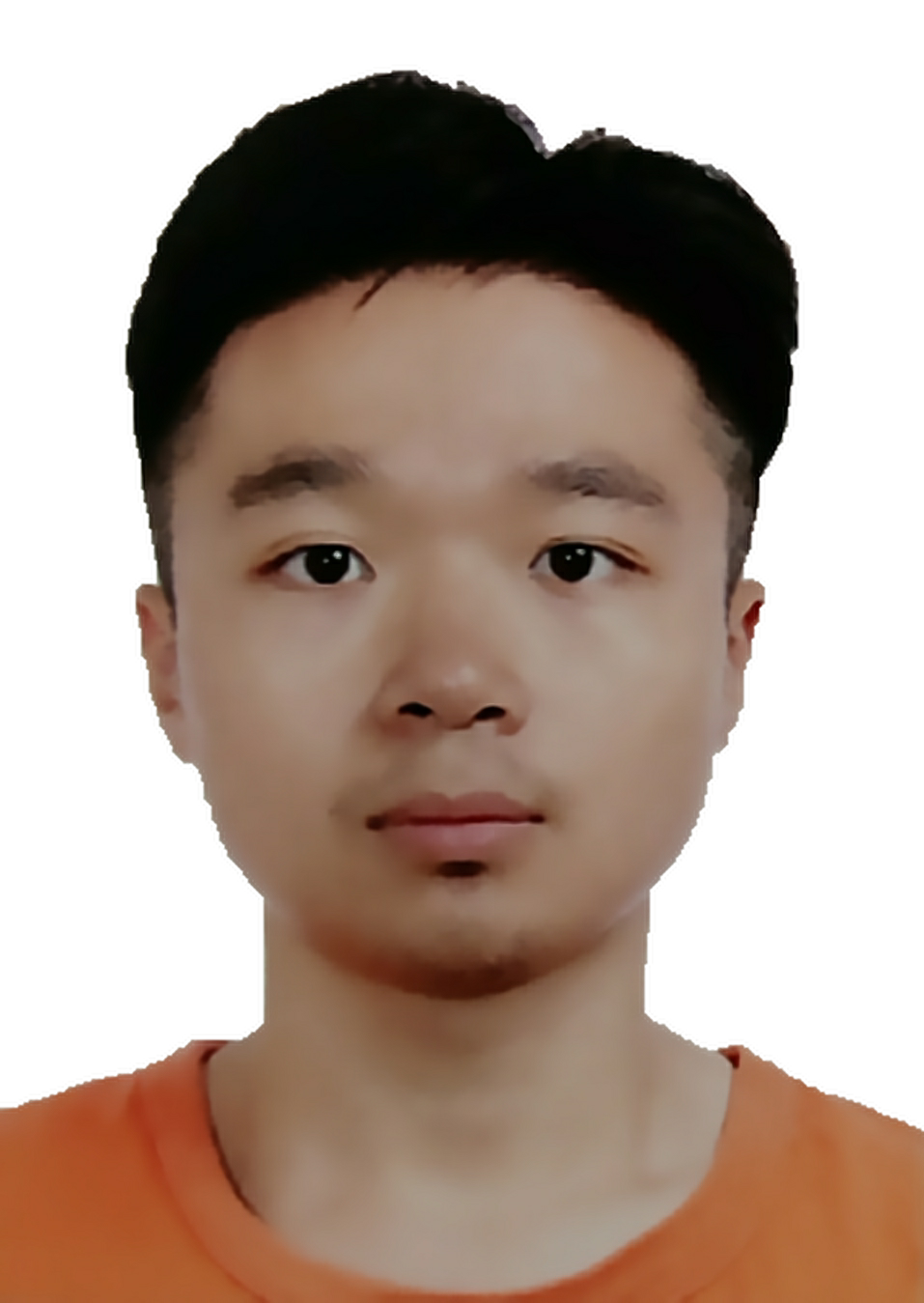}{Yuhuan Liu} is currently a Master student in the Department of Computer Science and Engineering at Southern University of Science and Technology (SUSTech). His research interests include indoor positioning and industrial control network security. He received his B.E. degree in 2019 with the major of electronic and information engineering from Nankai University, China.
\endbio

\bio[width=10mm,pos=l]{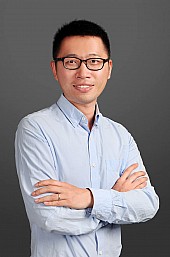}{Yulong Ding} received the B.A.Sc. and M.A.Sc. degrees in Chemical Engineering from Tsinghua University, China, in 2005 and 2008, respectively and the Ph.D. in Chemical Engineering from University of British Columbia, Canada, in 2012. He is currently a Research Assistant Professor with the Academy for Advanced Interdisciplinary Studies of Southern University of Science and Technology. His main interests are Industrial Internet of Things and Low-Power Wide Area Networks (LPWAN).
\endbio

\bio[width=10mm,pos=l]{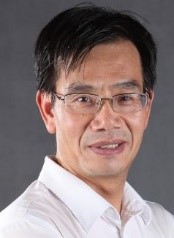}{Shuanghua Yang} received the B.S. degree in instrument and automation and the M.S. degree in process control from the China University of Petroleum (Huadong), Beijing, China, in 1983 and 1986, respectively, and the Ph.D. degree in intelligent systems from Zhejiang University, Hangzhou, China, in 1991. He was awarded DSc from Loughborough University in 2014 to recognize his academic contribution to wireless monitoring research. He is currently a chair professor of computer science and executive deputy dean of graduate school with Southern University of Science and Technology (SUSTech), Shenzhen, China.  Before joined SUSTech in 2016 he had spent over two decades in Loughborough University, as a professor in computer science and head of department. His current research interests include cyber-physical system safety and security, Internet of Things, wireless network-based monitoring and control. He is a Fellow of IET and a Fellow of InstMC, U.K. He is an Associate Editor of the IET Journal Cyber-Physical Systems Theory and applications, the InstMC Journal Measurement and Control, and the International Journal of Computing and Automation.
\endbio

\bio[width=10mm,pos=l]{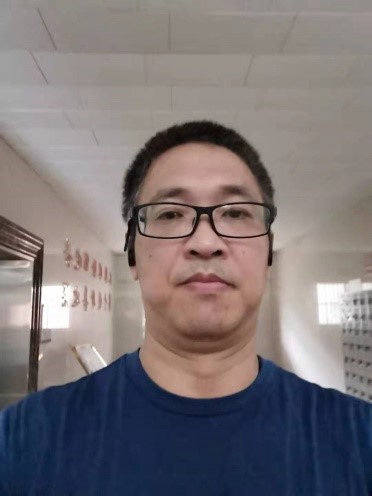}{Hao Li} received the M.S. degree in control engineering from The University of Science and Technology, China, in 2008. He is currently a professor of the Science and Technology on Near-surface Detection Laboratory of China. His main interests are Wireless sensor networks and Ad hoc network. 
\endbio

\end{document}